\documentclass[traditabstract, longauth]{aa}

\usepackage{natbib}
\usepackage{graphicx}
\usepackage{txfonts}
\usepackage{lipsum}

\usepackage{subcaption}         
\usepackage{lscape}             
\usepackage{placeins}           

\usepackage{orcidlink}
\usepackage{comment}
\usepackage{hyperref}

\newcommand{\bd}{\begin{displaymath}}
\newcommand{\ed}{\end{displaymath}}
\newcommand{\be}{\begin{equation}}
\newcommand{\ee}{\end{equation}}
\newcommand{\beaa}{\begin{eqnarray*}}
\newcommand{\eeaa}{\end{eqnarray*}}
\newcommand{\bea}{\begin{eqnarray}}
\newcommand{\eea}{\end{eqnarray}}

\newcommand{\zpertphot}{z_{\rm p,phot}}

\makeatletter
\let\linenumbers\relax
\makeatother

\begin{document}

   \title{HOLISMOKES XIX: SN~2025wny at $z=2$, the first strongly lensed superluminous supernova}

   \titlerunning{SN~2025wny, the first strongly lensed superluminous supernova}


\author{
Stefan~Taubenberger\inst{\ref{tum},\ref{mpa}}\orcidlink{0000-0002-4265-1958},
Ana~Acebron\inst{\ref{santander},\ref{iasf}}\orcidlink{0000-0003-3108-9039},
Raoul~Ca\~nameras\inst{\ref{lam}}\orcidlink{0000-0002-2468-5169},
Ting-Wan~Chen\inst{\ref{lulin}}\orcidlink{0000-0002-1066-6098},
Aymeric~Galan\inst{\ref{mpa},\ref{tum}}\orcidlink{0000-0003-2547-9815},
Claudio~Grillo\inst{\ref{milan},\ref{iasf}}\orcidlink{0000-0002-5926-7143},
Alejandra~Melo\inst{\ref{mpa},\ref{tum}}\orcidlink{0000-0002-6449-3970},
Stefan~Schuldt\inst{\ref{milan},\ref{iasf}}\orcidlink{0000-0003-2497-6334},
Allan~G.~Schweinfurth\inst{\ref{tum},\ref{mpa}}\orcidlink{0000-0002-8274-7196},
Sherry~H.~Suyu\inst{\ref{tum},\ref{mpa}}\orcidlink{0000-0001-5568-6052},
Greg~Aldering\inst{\ref{lbnl}},
Amar~Aryan\inst{\ref{lulin}}\orcidlink{0000-0002-9928-0369},
Yu-Hsing~Lee\inst{\ref{lulin}}\orcidlink{0009-0003-5139-9007},
Elias~Mamuzic\inst{\ref{mpa},\ref{tum}}\orcidlink{0009-0009-7962-656X},
Martin~Millon\inst{\ref{ethz},\ref{unige}}\orcidlink{0000-0001-7051-497X},
Thomas~M.~Reynolds\inst{\ref{turku},\ref{dawn},\ref{univcope}}\orcidlink{0000-0002-1022-6463},
Alexey~V.~Sergeyev\inst{\ref{oca},\ref{kha}}\orcidlink{0000-0003-3425-5178},
Ildar~M.~Asfandiyarov\inst{\ref{ulba}}\orcidlink{0009-0000-9605-9979},
Stéphane~Basa\inst{\ref{lam}},
Stéphane~Blondin\inst{\ref{esog},\ref{lam}}\orcidlink{0000-0002-9388-2932},
Otabek~A.~Burkhonov\inst{\ref{ulba}}\orcidlink{0000-0003-1169-6763},
Lise~Christensen\inst{\ref{univcope},\ref{dawn}}\orcidlink{0000-0001-8415-7547},
Frederic~Courbin\inst{\ref{iccub},\ref{icrea},\ref{ieec}}\orcidlink{0000-0003-0758-6510},
Shuhrat~A.~Ehgamberdiev\inst{\ref{ulba},\ref{ssu}}\orcidlink{0000-0001-9730-3769},
Tom~L.~Killestein\inst{\ref{warwick}}\orcidlink{0000-0002-0440-9597}, 
Seppo~Mattila\inst{\ref{turku},\ref{cyprus}},
Asadulla~M.~Shaymanov\inst{\ref{ulba}}\orcidlink{0009-0002-7815-5945},
Yiping~Shu\inst{\ref{pm}},
Dong~Xu\inst{\ref{altay}}\orcidlink{0000-0003-3257-9435},
Sheng~Yang\inst{\ref{henan}}\orcidlink{0000-0002-2898-6532},
Daniel~Gruen\inst{\ref{lmu},\ref{origins}}\orcidlink{0000-0003-3270-7644},
Justin~D.~R.~Pierel\inst{\ref{stsci},\ref{ef}}\orcidlink{0000-0002-2361-7201},
Christopher~J.~Storfer\inst{\ref{uh}}\orcidlink{0000-0002-0385-0014},
Kim-Vy~Tran\inst{\ref{cfa}}\orcidlink{0000-0001-9208-2143}
Kenneth~C.~Wong\inst{\ref{utokyo}}\orcidlink{0000-0002-8459-7793},
Rosa~L.~Becerra\inst{\ref{unam1}}\orcidlink{0000-0002-0216-3415},
Damien~Dornic\inst{\ref{cppm}}\orcidlink{0000-0001-5729-1468},
Jean-Grégoire~Ducoin\inst{\ref{cppm}},
Noémie~Globus\inst{\ref{unam2}}\orcidlink{0000-0001-9011-0737},
Claudia~P.~Guti\'errez\inst{\ref{ieec},\ref{ice}}\orcidlink{0000-0003-2375-2064},
Ji-an~Jiang\inst{\ref{ustchina},\ref{naoj}},
Hanindyo~Kuncarayakti\inst{\ref{turku}}\orcidlink{0000-0002-1132-1366},
Diego~López-Cámara\inst{\ref{unam3}}\orcidlink{0000-0001-9512-4177},
Peter~Lundqvist\inst{\ref{okc}},
Francesco~Magnani\inst{\ref{cppm}},
Enrique~Moreno~Méndez\inst{\ref{unam4}}\orcidlink{0000-0002-5411-9352},
Benjamin~Schneider\inst{\ref{lam}}\orcidlink{0000-0003-4876-7756},
Christian~Vogl\inst{\ref{mpa},\ref{tum}}
}

    \authorrunning{Taubenberger et al.}

    \institute{
    Technical University of Munich, TUM School of Natural Sciences, Physics Department, James-Franck-Str. 1, 85748 Garching, Germany \label{tum} 
    \and 
    Max Planck Institute for Astrophysics, Karl-Schwarzschild-Str. 1, 85748 Garching, Germany \label{mpa}
    \and
    Instituto de F\'isica de Cantabria (CSIC-UC), Avda.~Los Castros s/n, 39005 Santander, Spain \label{santander}
    \and
    INAF -- IASF Milano, via A. Corti 12, I-20133 Milano, Italy 
    \label{iasf}
    \and
    Aix-Marseille Université, CNRS, CNES, LAM, Marseille, France \label{lam}
    \and
    Graduate Institute of Astronomy, National Central University, 300 Jhongda Road, 32001 Jhongli, Taiwan \label{lulin}
    \and 
    Dipartimento di Fisica, Universit\`a  degli Studi di Milano, via Celoria 16, I-20133 Milano, Italy 
    \label{milan}
    \and
    E.O. Lawrence Berkeley National Laboratory, 1 Cyclotron Road, Berkeley, CA 94720, USA \label{lbnl}
    \and
    Institute for Particle Physics and Astrophysics, ETH Zurich, Wolfgang-Pauli-Str. 27, CH-8093 Zurich, Switzerland
    \label{ethz}
    \and
    D\'epartement de Physique Th\'eorique, Universit\'e de Gen\`eve, 24 quai Ernest-Ansermet, CH-1211 Gen\`eve 4, Switzerland
    \label{unige}
    \and
    Tuorla Observatory, Department of Physics and Astronomy, University of Turku, FI-20014 Turku, Finland \label{turku}
    \and
    Cosmic Dawn Center (DAWN), Niels Bohr Institute, University of Copenhagen, Jagtvej 128, DK-2200 Copenhagen N, Denmark \label{dawn}
    \and
    Universit\'e C{\^o}te d'Azur, Observatoire de la C{\^o}te d'Azur, CNRS, Laboratoire Lagrange, France
    \label{oca}
    \and
    V.N. Karazin Kharkiv National University, Kharkiv, Ukraine
    \label{kha}
    \and
    Ulugh Beg Astronomical Institute, 33 Astronomicheskaya St., Tashkent 100052, Uzbekistan
    \label{ulba}
    \and
    European Southern Observatory, Karl-Schwarzschild-Str. 2, 85748, Garching, Germany \label{esog}
    \and
    Niels Bohr Institute, University of Copenhagen, Jagtvej 128, DK-2200 Copenhagen N, Denmark\label{univcope}
    \and
    Institut de Ciències del Cosmos (ICCUB), Universitat de Barcelona (IEEC-UB), Martí i Franquès 1, 08028 Barcelona, Spain
    \label{iccub}
    \and
    Institució Catalana de Recerca i Estudis Avançats (ICREA), Passeig de Lluís Companys 23, 08010 Barcelona, Spain
    \label{icrea}
    \and
    Institut d’Estudis Espacials de Catalunya (IEEC), Edifici RDIT, Campus UPC, 08860 Castelldefels, Barcelona, Spain
    \label{ieec}
    \and
    Samarkand State University, 15, University boulevard, 140104, Samarkand, Uzbekistan
    \label{ssu}
    \and
    Department of Physics, University of Warwick, Gibbet Hill Road, Coventry CV4 7AL, UK \label{warwick}
    \and
    School of Sciences, European University Cyprus, Diogenes Street, Engomi, 1516, Nicosia, Cyprus \label{cyprus}
    \and
    Purple Mountain Observatory, Chinese Academy of Sciences, Nanjing, Jiangsu, 210023, China \label{pm}
    \and
    National Astronomical Observatories, Chinese Academy of Sciences, Beijing 100101, China \label{altay}
    \and
    Institute for Gravitational Wave Astronomy, Henan Academy of Sciences, Zhengzhou 450046, Henan, China \label{henan}
    \and
    University Observatory Munich, Faculty of Physics, Ludwig-Maximilians-Universität, Scheinerstr. 1, 81679 Munich, Germany
    \label{lmu}
    \and
    Excellence Cluster ORIGINS, Boltzmannstr. 2, 85748 Garching, Germany
    \label{origins}
    \and
    Space Telescope Science Institute, 3700 San Martin Drive, Baltimore, MD 21218, USA
    \label{stsci}
    \and
    NASA Einstein Fellow \label{ef}
    \and
    Institute for Astronomy, University of Hawaii, Honolulu, HI 96822-1897, USA \label{uh}
    \and
    Center for Astrophysics $|$ Harvard \& Smithsonian, 60 Garden Street, Cambridge, MA 02138, USA \label{cfa}
    \and
    Research Center for the Early Universe, Graduate School of Science, The University of Tokyo, 7-3-1 Hongo, Bunkyo-ku, Tokyo 113-0033, Japan \label{utokyo}
    \and
    Universidad Nacional Aut\'onoma de M\'exico, Instituto de Astronom\'ia, A.P. 70-264, 04510 Ciudad de M\'exico, M\'exico \label{unam1}
    \and
    Aix-Marseille Université, CNRS/IN2P3, Centre de Physique des Particules de Marseille (CPPM), IPhU, Marseille, France \label{cppm}
    \and
    Instituto de Astronom\'ia, Universidad Nacional Aut\'onoma de M\'exico, km 107 Carretera Tijuana-Ensenada, 22860 Ensenada, Baja California, México \label{unam2}
    \and
    Institute of Space Sciences (ICE, CSIC), Campus UAB, Carrer de Can Magrans, s/n, 08193 Barcelona, Spain \label{ice}
    \and
    Department of Astronomy, University of Science and Technology of China, Hefei 230026, China \label{ustchina}
    \and
    National Astronomical Observatory of Japan, 2-21-1 Osawa, Mitaka, Tokyo 181-8588, Japan \label{naoj}
    \and
    Instituto de Ciencias Nucleares, Universidad Nacional Autónoma de México, Apartado Postal 70-264, 04510 México, Ciudad de M\'exico, México \label{unam3}
    \and
    The Oskar Klein Centre, Department of Astronomy, Stockholm University, Albanova University Center, SE 106 91 Stockholm, Sweden\label{okc}
    \and
    Facultad de Ciencias, Universidad Nacional Autónoma de México, Apartado Postal 70-264, 04510 México, Ciudad de M\'exico, México\label{unam4} 
    }

   \date{Received XX XX, 2025}
 
  \abstract
   {We present imaging and spectroscopic observations of supernova SN~2025wny, associated with the lens candidate PS1~J0716$+$3821. Photometric monitoring from the Lulin and Maidanak observatories confirms multiple point-like images, consistent with SN~2025wny being strongly lensed by two foreground galaxies. Optical spectroscopy of the brightest image with the Nordic Optical Telescope and the University of Hawaii 88-inch Telescope allows us to determine the redshift to be $z_{\rm s} = 2.008 \pm 0.001$, based on narrow absorption lines originating in the interstellar medium of the supernova host galaxy. At this redshift, the spectra of SN~2025wny are consistent with those of superluminous supernovae of Type I. We find a high ejecta temperature and depressed spectral lines compared to other similar objects. We also measure, for the first time, the redshift of the fainter of the two lens galaxies (the ``perturber'') to be $z_{\rm p}=0.375\pm0.001$, fully consistent with the DESI spectroscopic redshift of the main deflector at $z_{\rm d}=0.3754$. 
   SN~2025wny thus represents the first confirmed galaxy-scale strongly lensed supernova with time delays likely in the range of days to weeks, as judged from the image separations. This makes SN~2025wny suitable for cosmography, offering a promising new system for independent measurements of the Hubble constant. Following a tradition in the field of strongly-lensed SNe, we give SN~2025wny the nickname SN Winny.
   }


   \maketitle

\section{Introduction}

The persistent tension in measurements of the Hubble constant, $H_0$, most notably between the local distance ladder using Type Ia supernovae (SNe) from SH0ES \citep[e.g.][]{2022riess} and early-Universe inferences from the Planck cosmic microwave background observations \citep{planck20}, highlights both gaps in our understanding of cosmology and the need for independent probes \citep[e.g.][]{moresco22, Verde+24}. Astrophysical transients, such as SNe, that are strongly lensed by foreground galaxies or galaxy clusters into multiple, time-delayed images offer a particularly powerful way to measure $H_0$ and other cosmological parameters \citep{1964Refsdal}. The time delays between the multiple transient images, together with a lens model of the total mass distribution, yield a direct measurement of the time-delay distance \citep{suyu10b}, and hence $H_0$, which is independent of both the distance ladder and early-Universe measurements \citep[see recent reviews by e.g.][]{Oguri19, Treu+22, Suyu+24}

Given the rarity of strongly lensed SNe, the time-delay method has been demonstrated through lensed quasars, which are more abundant \citep[e.g.][]{Suyu17, Wong20, Birrer20, millon20a, TDCOSMO2025}. The past decade has seen rapid progress in the discovery of strongly lensed supernovae following the first detection of SN~Refsdal \citep{Kelly15}. Since then, the number of known systems has grown to nearly ten; most of these SNe are lensed by galaxy clusters \citep[e.g.][]{rodney21, chen22, 2023Frye, 2024Pierel}, and only two systems are lensed by individual galaxies \citep{goobar17, 2023goobar}. Three of the galaxy-cluster systems have yielded the first $H_0$ measurements from lensed SNe \citep{Kelly23, grillo24, LiuOguri25, pascale+25, pierel25, suyu25, agrawal25}. The two previously known galaxy-scale systems, iPTF16geu and SN~Zwicky, had short time delays of $\lesssim 1$\,day \citep{dhawan20, pierel23}, which were not suitable for precision cosmography.  

With lensed SNe becoming an observational reality, we have initiated the HOLISMOKES programme \citep[Highly Optimised Lensing Investigations of Supernovae, Microlensing Objects,
and Kinematics of Ellipticals and Spirals;][]{Suyu+20}. The goals of the programme are to (1) measure the Hubble constant through the lensing time delays of SNe, and (2) constrain the properties of SN progenitors by using lensing time delays as a cosmic time machine, enabling early-phase observations of trailing SN images soon after their explosion.

Not only are lensed SNe an excellent probe of cosmology and SN progenitors, but the lensing effect also provides an amplified SN flux and a magnified view of their host galaxies. In fact, lensing clusters and galaxies have been used as nature's cosmic telescopes to study distant sources, which would be otherwise too faint to be observed with existing facilities \citep[e.g.][]{vanzella17, mestric23, messa25}. As an example, \citet{Dhawan+24} showed that SN Encore, a Type Ia SN at $z=1.949$, has a similar spectrum compared to those of local SNe, suggesting no cosmic evolution for the use of SNe Ia as a cosmological probe.

\begin{figure*}
    \centering
    \includegraphics[width=1.0\linewidth]{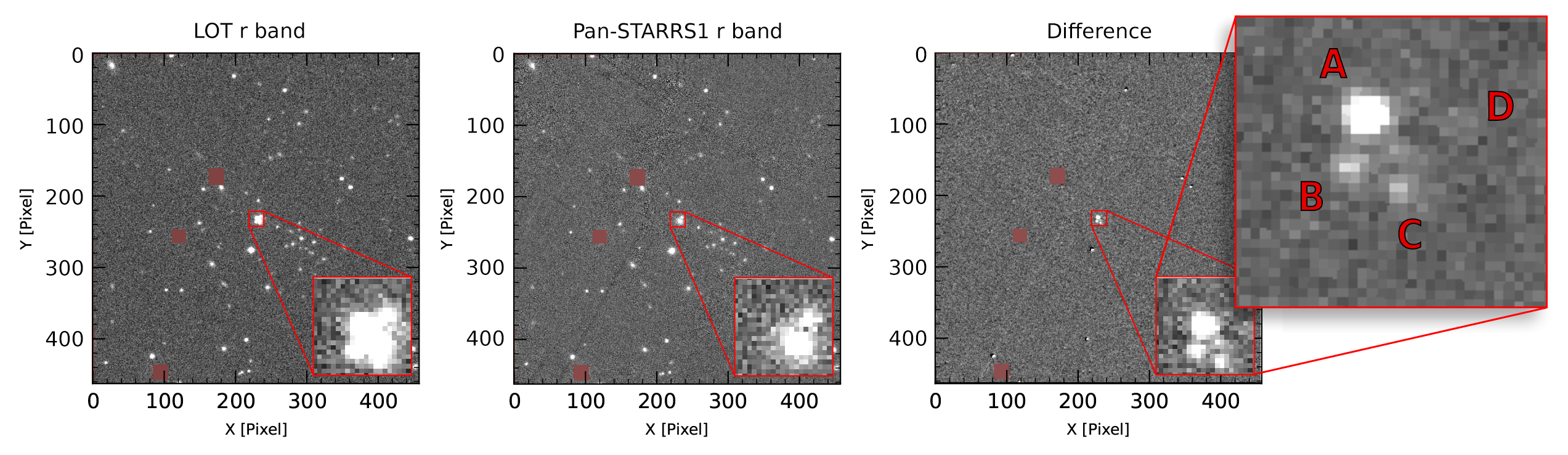}
    \caption{$r$-band image of SN~2025wny, obtained at the Lulin Observatory with LOT on September 29, 2025 (left), Pan-STARRS1 reference image (middle), and difference image after subtraction (right). Very bright stars have been masked. The images cover an area of roughly $9$\,arcmin$^2$. The SN images A to D are labelled on a zoomed-in view of the difference image.}
    \label{fig:LOT}
\end{figure*}

In this paper, we present a characterisation of SN~2025wny, the first galaxy-scale strongly lensed SN, which turns out to be a superluminous SN (SLSN). Its spatially resolved multiple images and time delays make it suitable for high-precision cosmography. In Sect.~\ref{sec:discovery}, we summarise the discovery and key properties of the supernova and lens system. Sect.~\ref{sec:imaging} describes follow-up imaging, and Sect.~\ref{sec:spectro} presents follow-up spectroscopy. We present the redshift determinations for SN~2025wny and the second lens galaxy, along with the SN classification, in Sect.~\ref{sec:results}. Finally, we present a summary and outlook in Sect.~\ref{sec:conclu}.


\section{SN~2025wny discovery}
\label{sec:discovery}

The Zwicky Transient Facility (ZTF) first detected SN~2025wny on August 27, 2025 (ID ZTF25abnjznp), but the Gravitational-wave Optical Transient Observer (GOTO) team first reported it to the Transient Name Server (TNS) on September 01, 2025 (ID GOTO25gtq). Based on Liverpool Telescope (LT) observations on UT October 3.21, 2025, the source was reported as a likely strongly lensed transient at RA\,=\,$07h\,16m\,34.500s$ and DEC\,=\,$+38d\,21m\,08.11s$, designated AT~2025wny \citep{2025TNSconf}.
Three point-like images were identified within $\sim$$2$\,arcsec from the strong lens candidate PS1~J0716$+$3821 (reported by our team, where we specifically searched for wide-separation lens systems as potential hosts of lensed transients that would be useful for cosmology; \citealt{canameras20}), and photometry of the brightest multiple image (A, see Fig.~\ref{fig:LOT}) yielded magnitudes of $g = 20.42 \pm 0.04$, $r = 19.60 \pm 0.03$, and $i = 19.54 \pm 0.03$ \citep{2025TNSconf}. 

\begin{figure}
    \centering
    \includegraphics[width=1.0\linewidth]{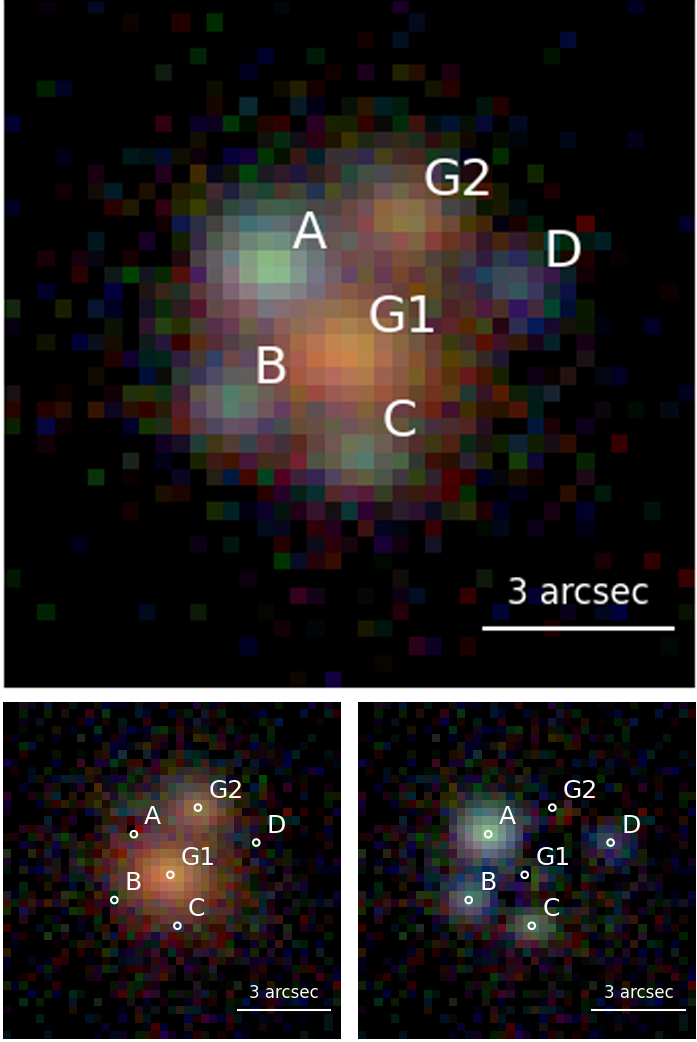}
    \caption{$VRI$-band colour composite of SN~2025wny obtained at the Maidanak Observatory. The four visible images of the SN are labelled by A, B, C and D (in order of decreasing brightness), and the two foreground deflector galaxies, by G1 and G2. \textit{Top:} observed image. \textit{Bottom left:} image of the deflectors after subtracting SN and host light using Moffat profiles (with joint shape parameters).  \textit{Bottom right:} image of SN after subtracting G1 and G2's light using a single Sérsic profile each.
    }
    \label{fig:maidanak_color}
\end{figure}

\begin{figure}
    \centering
    \includegraphics[width=1.0\linewidth]{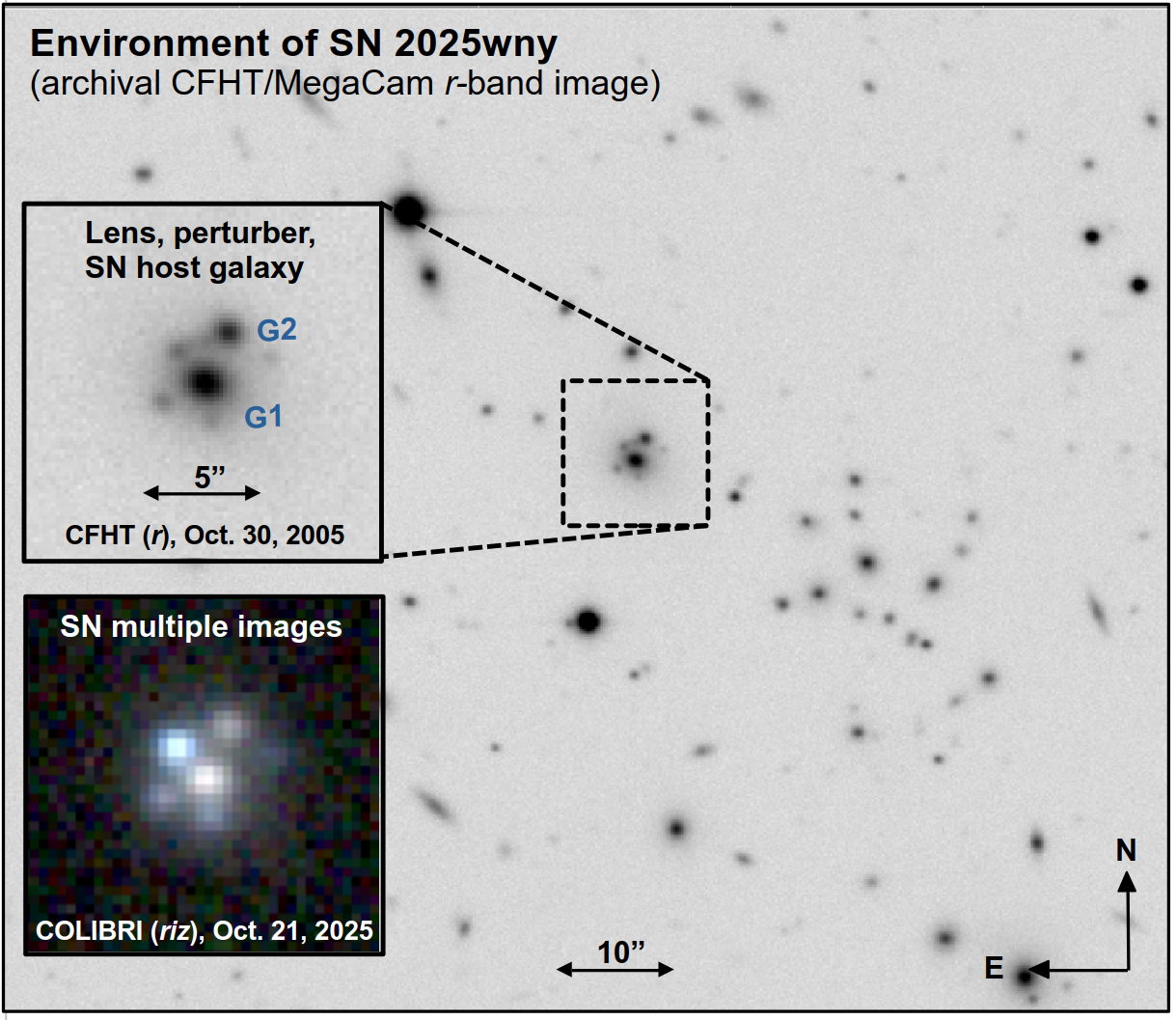}
    \caption{Archival CFHT image from 2005 \citep{2012Gwyn}, showing the two deflector galaxies G1 and G2 and four strongly lensed images of the SN host galaxy. A colour image generated from COLIBRI-telescope $riz$-band data is included to facilitate the comparison between the positions and brightnesses of the different SN and host-galaxy images.}
    \label{fig:cfht}
\end{figure}

The deflector is composed of a main lens and a smaller perturber galaxy, labelled as G1 and G2 in Fig.~\ref{fig:maidanak_color}, respectively. Wide-field imaging (from e.g. Pan-STARRS1, \citealt{2016chambers}; Legacy Surveys DR9, \citealt{2019Dey}; and see Fig.~\ref{fig:LOT}) shows that the lens system lies in a rich environment (see also Fig.~\ref{fig:cfht}), with tens of galaxies exhibiting similar colours and photometric redshifts. 
Archival data from the Canada-France-Hawaii Telescope (CFHT, \citealt{2012Gwyn}) from 2005 reveal four lensed images of the SN host galaxy in a cusp-like configuration (Fig.~\ref{fig:cfht}). The redshift of G1 is spectroscopically confirmed by the Dark Energy Spectroscopic Instrument (DESI; \citealt{2013Levi,2016DESIcollab}) at $z_{\rm d}=0.3754$, 
while G2 has a photometric redshift of $\zpertphot=0.32$ from the Dark Energy Camera Legacy Survey (DECaLS) DR9 imaging \citep{zhou23}. 

Our team identified the transient as a possible lensed SN by cross-matching our lens-candidate list with Lasair \citep{smith19} on September 21, 2025. Given its match with PS1~J0716+3821, SN 2025wny is the first detection of a wide-separation, galaxy-scale strongly lensed SN resulting from the search approach presented by \citet{canameras20}, which comprised (i) the systematic classification of all $3$ billion sources detected in the Pan-STARRS $3\pi$ survey with deep learning, (ii) the selection of high-quality strong-lens candidates, and (iii) their continuous positional cross-matching with the public ZTF alert stream. 
We present here our first follow-up imaging and spectroscopic data, obtained despite a series of technical and meteorological obstacles.


\section{Follow-up imaging}
\label{sec:imaging}

After identifying SN~2025wny as a promising lensed-SN candidate, we triggered follow-up imaging on September 29, 2025, at the Lulin Observatory (see Sect.~\ref{sec:lulin}) and on October 15, 2025, at the Maidanak Observatory (see Sect.~\ref{sec:maidanak}). In addition, we started monitoring with a two- to three-day cadence using the $1.3$\,m COLIBRI telescope \citep{basa22}\footnote{COLIBRI is a Franco-Mexican telescope at the Observatorio Nacional Astronómico San Pedro Mártir, Baja California, México, in operation since January 2025.} on October 15, 2025, in the $rz$-bands, which was later extended to the $riz$-bands. We also initiated $gri$-band observations with the $2.5$\,m Wide Field Survey Telescope \citep{WFST23} and the $1$\,m telescope at the Altay Observatory, with a few-day and daily cadence, respectively. Furthermore, we started imaging observations with the Three Channel Imager (3KK; \citealt{lang2016wendelstein}) mounted on the Fraunhofer Telescope at Wendelstein Observatory (FTW; \citealt{2014SPIE.9145E..2DH}).  We provide a brief overview of the first imaging data below, while full details will be presented in a forthcoming publication once the imaging campaigns are completed. At that time, we will also provide all the magnitude measurements, the full light curves, and the measured time delays between the different strongly lensed images.

\subsection{Lulin Observatory}
\label{sec:lulin}

We began our first imaging campaign on September 29, 2025, with the Lulin One-meter Telescope (LOT) at the Lulin Observatory as part of the DETECT (DESI Transient Event Cross-matching Tool) program (Y.-H.~Lee et al., in prep.). 
Since then, we have obtained multiple epochs of $gri$-band imaging. 
The images were reduced following a standard procedure using a customised pipeline\footnote{https://hdl.handle.net/11296/98q6x4}, including bias subtraction, dark correction, and flat-fielding. 
Figure~\ref{fig:LOT} shows the first $r$-band image taken on September 29, 2025, obtained under very good observing conditions. We used AutoPhOT \citep{2022A&A...667A..62B} to perform the template subtraction. A difference image between this new observation and a Pan-STARRS1 image, serving as a template image without the lensed transient, clearly reveals the three transient images A, B, and C, and a marginal detection of image D (see Fig.~\ref{fig:LOT}, right panel). Therefore, this observation confirms that SN~2025wny is indeed strongly lensed.  We measure $r$-band AB magnitudes of $19.6$\,mag, $21.3$\,mag, $21.5$\,mag for images A, B, and C, respectively.

\subsection{Maidanak Observatory} 
\label{sec:maidanak}

Starting from October 15, 2025, we expanded our monitoring campaign by using the $1.5$\,m telescope of the Maidanak Observatory (Uzbekistan) \citep{2018NatAs}. We obtained imaging data in the $V$, $R$, and $I$ filters \citep{2010JKAS}, each with about $30$\,min exposure time. 
Figure~\ref{fig:maidanak_color} (top) shows a colour composite of our first epoch of $VRI$ data, obtained under good seeing conditions (Full Width at Half Maximum, ${\rm FWHM_{V}}=0.96$\,arcsec$,\; {\rm FWHM_{R}}=0.72$\,arcsec$,\; {\rm FWHM_{I}}=0.83$\,arcsec). In the bottom-left panel of Fig.~\ref{fig:maidanak_color}, we show the light of the lens galaxies G1 and G2, and on the bottom-right panel the SN (+\,host) images only.

We measure $V$, $R$ and $I$-band Vega magnitudes of $20.31\pm0.02$\,mag, $19.48\pm0.01$\,mag and $19.17\pm0.01$\,mag for image A using Point Spread Function (PSF) photometry with Moffat functions \citep{1969A&A.....3..455M}. 
The reference star used for the photometric calibration is at RA\,=\,$07h\,16m\,34.78s$ DEC\,=\,$+38d\,20m\,51.0s$, with magnitudes of $G=17.781$, $BP=18.538$, $RP=16.858$ from Gaia DR3 \citep{2021A&A...649A...1G}. We transformed the Gaia magnitudes into the Johnson-Cousins system using the equations from \citet{2021A&A...649A...3R}.
The uncertainties were estimated based on the background noise and the PSF fitting residuals. 
The reported magnitudes include some contamination from the underlying SN host galaxy, but given the brightness of the SN image A relative to the host, the host contribution is at most a few percent. Difference-imaging magnitudes as well as
magnitudes for the remaining images and for galaxies G1 and G2 will be presented in a forthcoming paper.

\section{Spectroscopic observations with NOT and UH88}
\label{sec:spectro}

Optical spectroscopy of SN~2025wny image A was obtained with the Nordic Optical Telescope (NOT) + ALFOSC as part of the NOT Un-biased Transient Survey 2 (NUTS2\footnote{\url{https://nuts.sn.ie/}}) programme on UT October 11.16, 2025. Two exposures of $1800$\,s each were taken with grism 4 and the $1.0$\,arcsec slit, resulting in a wavelength range from $3300$ to $9700\,\text{\AA}$ and a spectral resolution $\rm R \sim 350$. The slit was aligned along the parallactic angle, which serendipitously included part of the perturber galaxy (G2 in Fig.~\ref{fig:maidanak_color}), for which no spectroscopic redshift was known before. 

The data were reduced following standard procedures in IRAF\footnote{NOIRLab IRAF is distributed by the Community Science and Data Center at NSF NOIRLab, which is managed by the Association of Universities for Research in Astronomy (AURA) under a cooperative agreement with the U.S. National Science Foundation.}, followed by an optimal, variance-weighted extraction of the spectra with the IRAF task \texttt{apall}. Wavelength calibration was accomplished with a HeNe arc-lamp spectrum and checked against night-sky emission lines. A spectrum of the spectrophotometric standard star BD+17d4708, obtained during the same night with identical setup, was used for flux calibration and telluric correction. We checked the flux calibration against the Maidanak photometric observations of October 15 (Sect.~\ref{sec:maidanak}), and found that the synthetic colours match the observed ones at the $2$-percent level.

A second spectrum with the NOT + ALFOSC was obtained on UT October 19.18, 2025 (thanks to contributed programmes 68-804 and 71-804). This time, grism 18 was used in combination with the $1.0$\,arcsec slit to achieve a higher resolution of $\rm R \sim 1000$ for the blue part of the spectrum ($3450-5350\,\text{\AA}$). The slit was aligned to include images A and C, and the exposure time was $2 \, \times \, 2400$ s. The data reduction was carried out in the same way as for grism 4, with the standard star He 3 used for flux calibration.

Additionally, Integral Field Spectroscopy (IFS) of the SN~2025wny system was obtained on UT October 15.59, 2025, with the University of Hawaii 88-inch Telescope (UH88) and the SuperNova Integral Field Spectrograph (SNIFS; \citealt{lantz2004}). SNIFS has a microlens array of $15 \times 15$ spaxels, each $0.43 \times 0.43$\,arcsec$^2$, giving a total field of view $6.4 \times 6.4$\,arcsec$^2$. This covers the entire SN~2025wny system, so the exposures contain the four images and the two lensing galaxies. The light is divided into blue (B) and red (R) channels, nominally spanning $3300-5150\,\text{\AA}$ and $5100-9700\,\text{\AA}$, respectively. The corresponding spectral resolutions are $5.2\,\text{\AA}$ and $7.2\,\text{\AA}$. Four exposures of $30$\,min each were obtained, during which the seeing was in the range $1.1-1.3$\,arcsec FWHM (somewhat worse than usual due to a warm primary mirror following an extended telescope shutdown). The combination of passing cirrus and moonlight further lowered the signal-to-noise ratio (SNR).

\begin{figure*}
    \centering
    \includegraphics[width=0.9\linewidth]{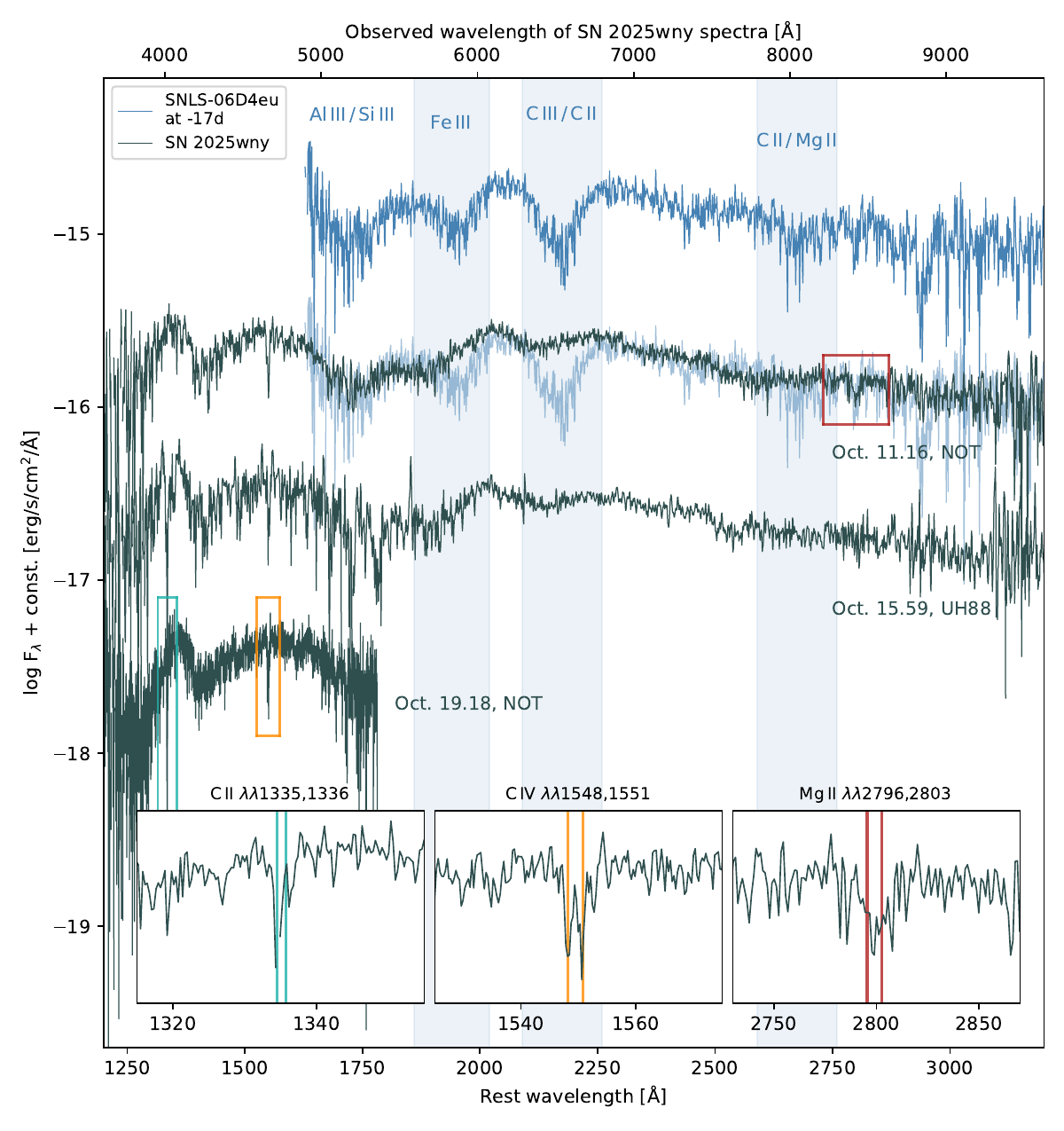}
    \caption{Spectra of SN~2025wny, obtained with NOT + ALFOSC and UH88 + SNIFS. A spectrum of the SLSN-I SNLS-06D4eu from \citet{howell2013} is included for comparison, plotted once with and once without offset relative to the SN~2025wny spectrum of October 11.16. SNLS-06D4eu provides the best match with SN~2025wny. Line identifications for SNLS-06D4eu have been adopted from \citet{howell2013} and \citet{mazzali2016}, but we note that in SN~2025wny, the absorptions marked by the blue-shaded bands are weaker and more strongly blueshifted. The inserts at the bottom zoom in onto narrow absorption lines from the ISM in the host of SN~2025wny. They have been used to determine the redshift of the SN to be $z_{\rm SN} = 2.008 \pm 0.001$.}
    \label{fig:seq}
\end{figure*}

The SNIFS pipeline takes associated arcs and flats and constructs a 3D data cube of $\alpha, \delta, \lambda$. Whereas PSF photometry is usually used to extract SNe after subtracting a reference datacube obtained a year or more later \citep{bongard2011}, that was not possible at this stage, so aperture extraction was used instead. To optimise the SNR and minimise the impact of position errors, while avoiding contamination from the lensing galaxy and other sources, a $2$\,arcsec diameter aperture (i.e. a radius of $\sim$$0.8$\,FWHM) was used \citep{howell1989, king2013}. The aperture locations were set separately for the B and R channels, but were not corrected for atmospheric differential refraction as a function of wavelength within a channel. The maximum effect of this simplification is well below the noise, and can be removed altogether in a future analysis that includes a reference datacube or model for the lensing galaxy. With each spectral extraction, the variance spectrum is estimated from the photon statistics and detector noise. As the SNR was similar for all four spectra, they were summed to produce the spectrum shown in Fig.~\ref{fig:seq}. Flux calibration and telluric corrections used the pipeline defaults \citep{buton2013}.
Here, we show the spectral extraction only for image A, as the others yielded much lower SNR, and also had stronger contamination from the lensing galaxy. 

All three spectra of image A are shown in Fig.~\ref{fig:seq}.

\section{Results}   
\label{sec:results}

\subsection{The redshift of SN~2025wny}
\label{sec:redshift}

Measuring the redshift of SN~2025wny turned out to be challenging at first, since the broad, blended spectral features in the blue part and the almost featureless continuum above $\sim$$7000\,\text{\AA}$ (observer frame) in our NOT spectrum of October 11.16 did not immediately tick any boxes (see Fig.~\ref{fig:seq}). The only constraint from strong lensing with a deflector at $z_\mathrm{d} = 0.375$ was that the redshift of the transient was most likely $\gtrsim 0.5$. Various SN classification codes (SNID, \citealt{blondin2007}; GELATO, \citealt{harutyunyan2008}; NGSF, \citealt{howell2005,goldwasser2022}) were used to classify the transient, but they all failed to provide compelling matches due to the lack of comprehensive spectral databases of SNe in the rest-frame UV.

The key to the SN redshift, therefore, could only come from narrow absorptions in the interstellar medium of the SN host galaxy. The most prominent narrow absorption in our NOT spectrum of October 11.16 is located at $4663\,\text{\AA}$, and while it is not resolved, its FWHM is larger than expected for a single line. The most obvious guess for the nature of this line was Mg\,\textsc{ii} $\lambda\lambda2796,2803$, which would have set the SN at a redshift $z = 0.665$. To verify or reject this hypothesis, we obtained a higher-resolution spectrum on October 19.18, in which the feature is clearly separated into two absorptions. The separation, however, measured to be $\sim$$7\,\text{\AA}$, is smaller than the  $\sim$$12\,\text{\AA}$ expected for Mg\,\textsc{ii}. A convincing match was finally found for C\,\textsc{iv} $\lambda\lambda1548,1551$ at $z_\mathrm{SN} = 2.008 \pm 0.001$, which is further supported by the detection of C\,\textsc{ii} $\lambda\lambda1335,1336$ and unresolved Mg\,\textsc{ii} $\lambda\lambda2796,2803$ at the same redshift (see Fig.~\ref{fig:seq}, bottom panels).

\subsection{The redshift of the perturber galaxy}
\label{sec:redshift_perturber}

\begin{figure}
    \centering
    \includegraphics[width=1.0\linewidth]{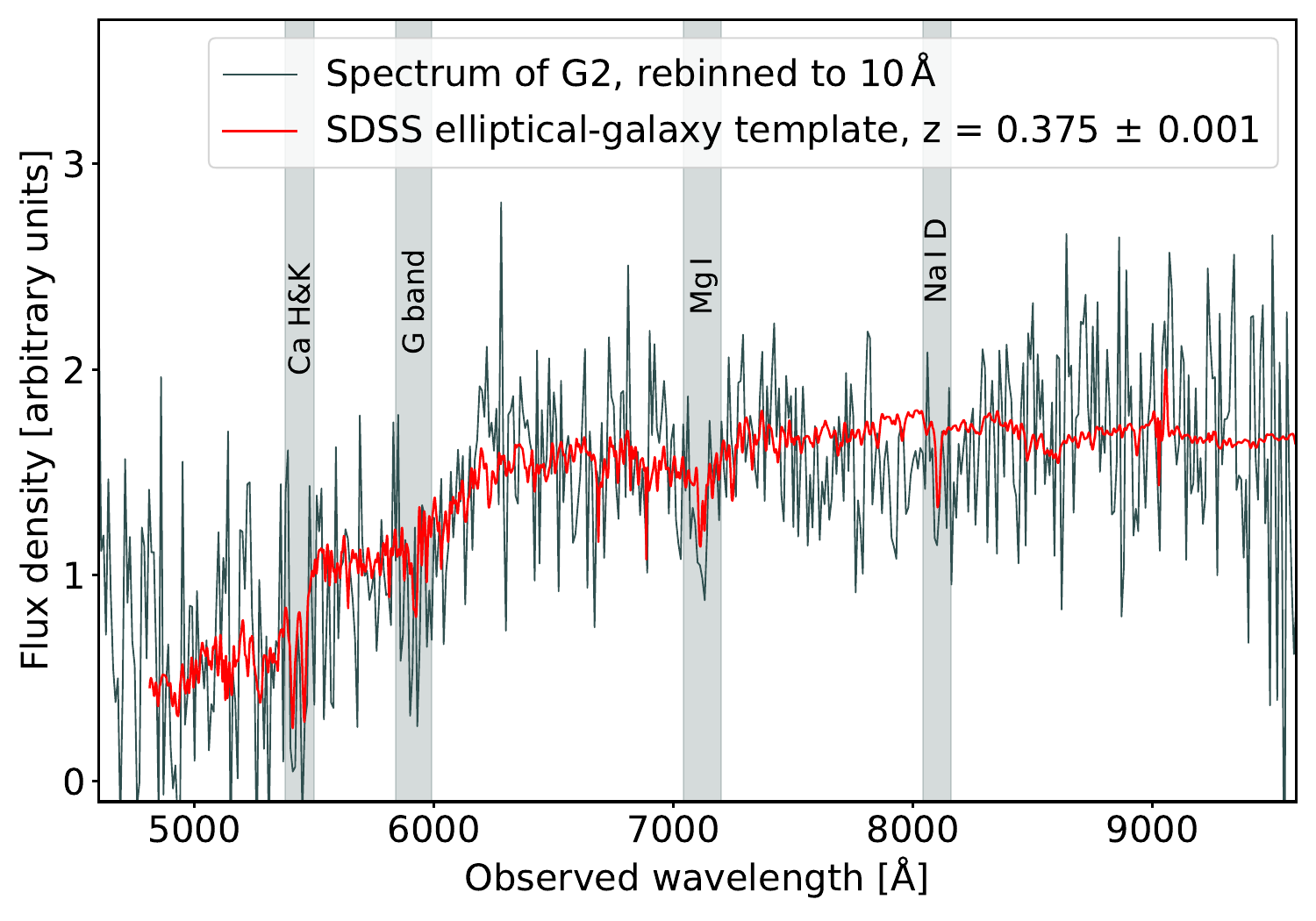}
    \caption{Spectrum of the second deflector galaxy G2 (`perturber'), rebinned to a $10\,\text{\AA}$ bin size. An SDSS DR5 template spectrum of an early-type galaxy, shifted to a redshift $z_\mathrm{p} = 0.375 \pm 0.001$, is superimposed. Ca\,\textsc{ii} H\&K, the Balmer break, the G band, Mg\,\textsc{i}, and Na\,\textsc{i} D are clearly detected.}
    \label{fig:perturber}
\end{figure}

Not only the redshift of the strongly lensed SN, but also that of the second deflector galaxy G2 (the perturber) was previously unknown. Until now, only a photometric redshift estimate of $\zpertphot = 0.32$ from DECaLS was available. However, accurately knowing the redshift of G2 is crucial for determining its effect on multiple image positions, lensing magnifications, and time delays. The fact that G2 was also included in the slit in the October 11.16 NOT spectrum allowed us to determine, for the first time, its spectroscopic redshift. Unfortunately, G2 is faint and was not perfectly centred in the slit, resulting in a low SNR of the extracted spectrum. Yet, the most important features of a passive-galaxy spectrum (Ca\,\textsc{ii} H\&K, Balmer break, G band, Mg\,\textsc{i}, Na\,\textsc{i} D) can be clearly identified. Fig.~\ref{fig:perturber} shows the NOT spectrum with a Sloan Digital Sky Survey (SDSS) Data Release 5 \citep{SDSS5} template spectrum\footnote{https://classic.sdss.org/dr5/algorithms/spectemplates/} of an early-type galaxy overlaid. From this match, we infer a redshift $z_\mathrm{p} = 0.375 \pm 0.001$ for the perturber G2, fully consistent with the spectroscopic redshift of the main lens G1. Hence, these two galaxies likely form a physical system.

\subsection{A particularly UV-bright superluminous SN}
\label{sec:SLSN}

\begin{figure}
    \centering
    \includegraphics[width=1.0\linewidth]{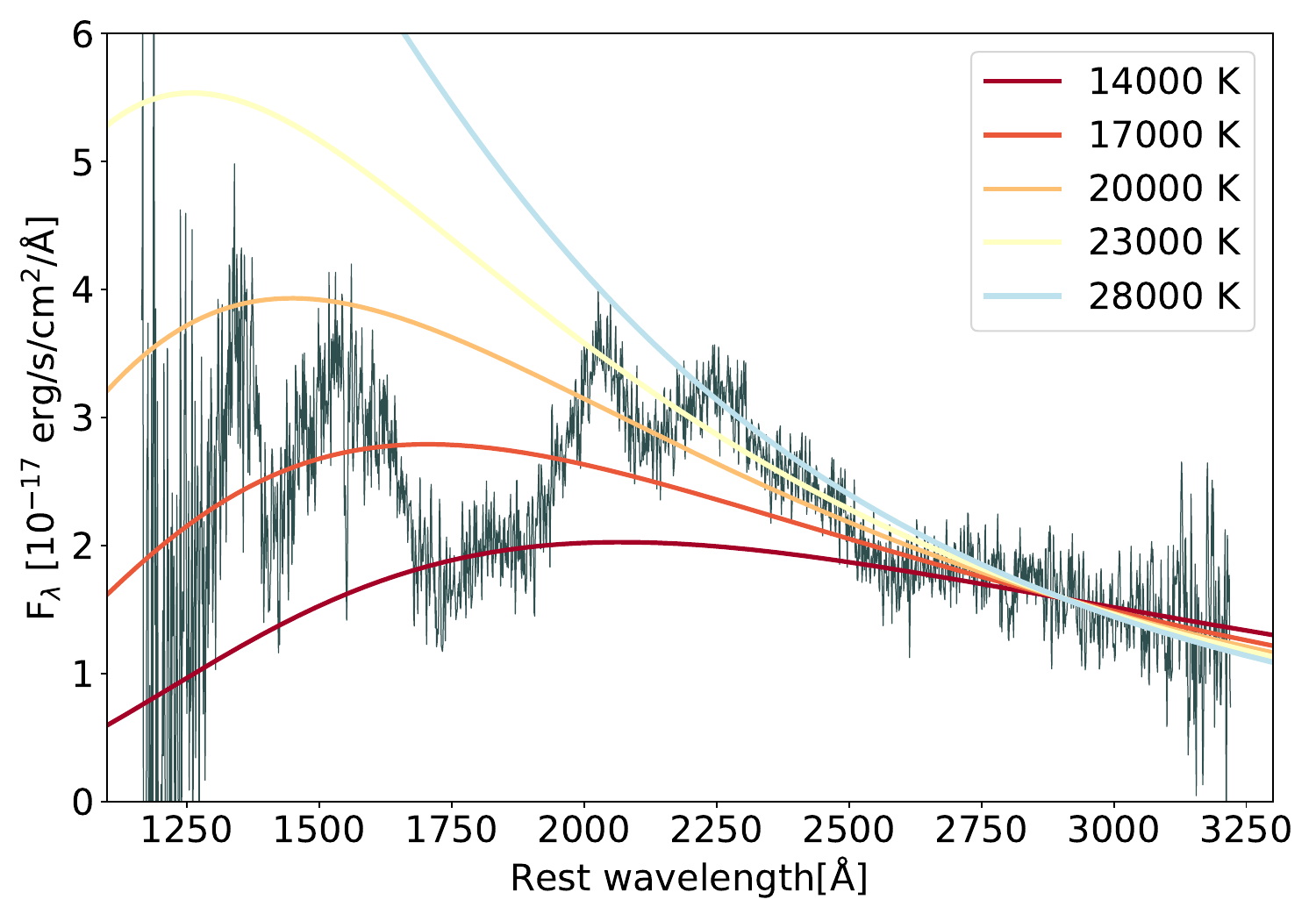}
    \caption{Spectrum of SN~2025wny of October 11.16 with blackbody spectra for different temperatures overlaid. The spectra have been normalised at $2900\,\text{\AA}$. While an exact determination of the ejecta temperature is hampered by the strong and broad spectral features blueward of $2000\,\text{\AA}$, temperatures $\gtrsim17\,000$\,K are favoured.}
    \label{fig:bb}
\end{figure}

Having the redshift established, the quest for the true identity of SN~2025wny continued. At $z_\mathrm{SN} = 2.008 \pm 0.001$, the entire observed optical spectrum probes the rest-frame UV. Clearly, there is no strong UV flux suppression or line blanketing as for most conventional SN types, especially SNe~Ia. Instead, the Spectral Energy Distribution (SED) peaks between $1300$ and $2300\,\text{\AA}$ (rest-frame wavelength). A comparison of the October 11.16 spectrum to blackbody curves (Fig.~\ref{fig:bb}) suggests very high ejecta temperatures of at least $\sim$$17\,000$\,K. Moreover, since the spectrum was taken $45$\,observer-frame days after discovery, corresponding to $15$\,rest-frame days, the high temperatures and UV flux were not just observed for a brief moment shortly after the explosion, but persisted for several weeks.

The only class of SNe consistent with these characteristics are SLSNe, especially SLSNe-I. These objects, which were first identified as a distinct class by \citeauthor{quimby2011} in 2011, are usually explained by the explosion of very massive stars or the formation and subsequent spin-down of a magnetar \citep[e.g.,][]{inserra2013}. They are UV-bright during their entire rise and around peak, a phase that often lasts for about $30-50$\,days \citep[e.g.][]{gomez2024}. Below $2000\,\text{\AA}$, they are typically one to two orders of magnitude more luminous than other SN types, even when normalised to the same flux at optical wavelengths \citep[see, e.g. Fig.~8 of][]{yan2017}. Their peak absolute magnitudes at rest-frame optical wavelengths typically range from $-20$ to $-22.5$ Mag \citep{gomez2024}, which makes them observable even at large distances. Accordingly, photometrically typed SLSNe-I have been observed out to a redshift $z \sim 4$ \citep{cooke2012}, whereas for spectroscopically confirmed events, the record holders are DES16C2nm at a redshift $z = 1.998$ \citep{smith2018} with several high-SNR spectra, and HSC16adga at $z = 2.399$ \citep{curtin2019}, whose redshift and classification are on more shaky grounds given the extremely low SNR. With our spectroscopically confirmed redshift of $2.008 \pm 0.001$, SN~2025wny is therefore among the most distant spectroscopically confirmed SLSNe, making this a showcase example of how strong lensing can help study transients that would otherwise be mostly out of reach.

SN~2025wny is somewhat special among SLSNe-I in featuring an unusually smooth UV spectrum. The part between $2200$ and $3200\,\text{\AA}$ is particularly featureless -- a region where most SLSNe-I show several prominent, broad spectral lines \citep[e.g.,][]{howell2013,yan2017}. However, as shown in Fig.~\ref{fig:seq}, we do find a convincing spectral match with the SLSN-I SNLS-06D4eu \citep{howell2013}, whose spectral features redward of $2200\,\text{\AA}$ are also strongly depressed. With a peak absolute magnitude $M_U = -22.7$ and its high UV flux, SNLS-06D4eu was quite extreme even by SLSN standards. \citet{howell2013} estimated a blackbody temperature of $13\,500-14\,000$\,K at the time the spectrum was taken ($17$\,rest-frame days prior to the $U$-band peak), implying an SED peak around $2100\,\text{\AA}$. However, this may actually be an underestimate: in the wavelength region in common, the continuum in SNLS-06D4eu and SN~2025wny overlaps very well (Fig.~\ref{fig:seq}), but the SN~2025wny spectrum reaches further into the UV thanks to its higher redshift and shows significant emission below $1700\,\text{\AA}$, requiring blackbody temperatures $\gtrsim 17\,000$\,K to be properly reproduced (Fig.~\ref{fig:bb}). A higher temperature is also supported by spectral modelling of SNLS-06D4eu by \citet{mazzali2016}, who infer a blackbody temperature of 18\,000\,K, the highest in their sample of SLSNe-I.

The main spectral differences between SNLS-06D4eu and SN~2025wny concern the strengths and positions of three spectral lines near $1900$, $2200$, and $2700\,\text{\AA}$ (marked by blue-shaded bands in Fig.~\ref{fig:seq}). These lines are significantly shallower and more strongly blueshifted in SN~2025wny. \citet{howell2013} employed SEDONA radiative-transfer calculations \citep{kasen2006} of chemically homogeneous ejecta to identify these features as blends of mostly carbon lines, with contributions from magnesium and iron. This identification was largely confirmed by \citet{mazzali2016}, who, however, found a smaller contribution from singly ionised carbon due to the high temperature. \citet{mazzali2016} explained the feature near 1700\,\AA\ by resonance lines of doubly ionised aluminium and silicon. This feature is strong regardless of the overall ejecta composition, since it already forms at a solar metal abundance \citep{howell2013}. 
\citet{howell2013} showed that the entire UV spectrum of SNLS-06D4eu can be well reproduced with a composition that is a mix of carbon and oxygen, with a solar admixture of heavier elements. Helium-dominated ejecta do not show the prominent carbon features and lead to an overall smoother appearance, since all helium lines in the respective wavelength range turn out to be weak at the given density and temperature \citep[see Fig.~12 of][]{howell2013}.

This finding might actually be a key to understanding the differences between SNLS-06D4eu and SN~2025wny: if the ejecta in SN~2025wny were more helium-rich, with a lower abundance of carbon and oxygen, the weakness of the carbon lines could be explained naturally. Alternatively, the differences in the strength of particular lines and their position could also arise from higher ejecta velocities in SN~2025wny or from a phase mismatch, since we do not yet have sufficiently good photometric coverage to determine the phase of SN~2025wny's image A.

\subsection{The odds of finding a strongly lensed superluminous SN}
\label{sec:odds}

In the local Universe, superluminous SNe are extremely rare, but \citet{Prajs2017} showed that their rate increases with redshift, consistent with the finding that their host galaxies are predominantly metal-poor (e.g., \citealt{2013ApJ...763L..28C}). However, even at $z\sim1$, their rate is only a few $\times\,10^{-4}$ of the volumetric rate of core-collapse SNe at the same redshift \citep{Prajs2017,2021MNRAS.500.5142F}. 
Given this extreme rareness, one might wonder whether the detection of a strongly lensed SLSN is just a remarkable coincidence (reminiscent of the first gravitational-wave detection of a neutron-star merger and its electromagnetic counterpart; \citealt{abbott2017}) or whether one might expect to find them more regularly in strong-lensing systems in the future. Without going into any detailed rate calculations, which are beyond the scope of this paper, we want to lay out a few arguments why strongly lensed SLSNe might be discovered more frequently than previously thought, despite their low intrinsic rate:
\begin{itemize}
    \item SLSNe are by far the most luminous SNe. Even at rest-frame-optical wavelength, they are about $2$\,mag more luminous than typical SNe~Ia.
    \item They are also extremely UV-bright before and around peak. Their UV--optical colours are $3-4$\,mag bluer than those of SNe~Ia \citep[see, e.g. Fig.~8 of][]{yan2017}. At $z\gtrsim 1.5$, the observed optical light corresponds to the rest-frame UV regime. Hence, optical transient surveys such as ZTF and the Rubin Observatory Legacy Survey of Space and Time (LSST) are biased to detect UV-bright transients at those redshifts.
    \item If we make a thought experiment and replace SN~2025wny by a normal SN~Ia at the same location and redshift, its observed $r$-band peak would be $5-6$\,mag fainter than that of SN~2025wny. Hence, even the highly magnified image A would peak around $25$\,mag, images B to D closer to $27$\,mag. This is much fainter than the $5\sigma$ detection limit of LSST in single frames ($\sim$23--24\,mag, depending on the band). 
    \item SLSNe have intrinsically slowly evolving light curves, which are stretched out even more due to time dilation at high redshift. It is therefore very unlikely to miss a strongly lensed SLSN because of bad weather, technical failures, or even the seasonal gap.
    \item Little is known about the rate of SLSNe beyond redshift $1$, but extrapolating the trend from lower-redshift rate studies, they might arguably be more common there. 
\end{itemize}

Present simulations of lensed-SN rates and detection prospects in ongoing and future transient surveys \citep{oguri2010,wojtak2019,Arendse+24,Bag+24} do not consider SLSNe as a class, probably due to their low volumetric rate. Following SN~2025wny's discovery, repeating such calculations for SLSNe would certainly be warranted.

\section{Summary and outlook}
\label{sec:conclu}

We have presented the first characterisation of SN~2025wny, the first confirmed galaxy-scale strongly lensed supernova with expected time delays of days to weeks. The system shows four images produced by a two-galaxy deflector (G1\,+\,G2), and the transient spectrum is consistent with a SLSN at $z_\mathrm{SN} = 2.008 \pm 0.001$. The fact that a SLSN, despite being intrinsically rare among all types of SNe, is one of the first galaxy-scale lensed SN systems to be detected is likely due to selection effects; given the current limiting depth of the ZTF survey, only the brightest SNe are expected to be spatially resolved and detected when strongly lensed.

We have also presented key information on the G1\,+\,G2 system: measuring a spectroscopic redshift for G2 ($z_\mathrm{p} = 0.375 \pm 0.001$) helps to understand its impact on image positions, magnifications, and time delays, and thus provides critical input for modelling this complex lens system.	

Immediate priorities are continued high-cadence photometric monitoring to measure the time delays between the multiple SN images. Ongoing observations with Maidanak, Lulin, COLIBRI and Wendelstein will deliver the light curves required for time-delay measurements. In addition, spectral monitoring of the evolution of the multiple SN images can also be used to determine the time delays \citep[e.g.,][]{Bayer21, Johansson21}. High-resolution imaging of this system from adaptive-optics-assisted ground-based facilities or space-based observatories will also be important for detailed lens mass modeling. Both Hubble Space Telescope (Programme ID 17611; PI: Goobar) and James Webb Space Telescope (Programme ID 5564; PI: Goobar) were triggered to obtain follow-up imaging and IFU spectroscopy \citep{2025TNSconf}. The combination of time delays, lens mass modelling and lens environment analysis has the potential to provide an independent and competitive determination of the value of $H_0$.  

We are entering a new and exciting era with the discovery of this first galaxy-scale lensed SN system suitable for cosmography.  The discovery demonstrates the effectiveness of our lensed SN search strategy in HOLISMOKES of identifying static strong-lens systems in imaging surveys, cross-matching them to transient alerts, and flagging the crossmatches as lensed SN candidates \citep{canameras20}. With the imminent start of the LSST, we anticipate $\sim$$10$ lensed SNe Ia per year that are useful for cosmography \citep{Arendse+24}, and the combination of Euclid and LSST will be an effective way to find lensed SNe \citep{SainzdeMurieta+24}.  A sample of $\sim$$20$ lensed SNe Ia will not only enable us to rigorously test the time-delay methodology applied to lensed quasars, but will also yield a measurement of $H_0$ with $1$-percent uncertainty, which is crucial for resolving the Hubble tension \citep[e.g.][]{Suyu+24}.

For future reference, we propose to name this interesting object SN Winny to make it more memorable. The name Winny arises from its designation, SN~2025wny, and evokes warmth and companionship - a gentle light in the dark. Given its distant, radiant glow, we deem this name very appropriate.\\[-0.1cm]

\textbf{Note:} During the final preparation of this manuscript, a TNS AstroNote \citep{johansson2025} and classification report (\href{https://www.wis-tns.org/object/2025wny/classification-cert}{https://www.wis-tns.org/object/2025wny/classification-cert}) were published, independently confirming our results on both the redshift and classification of SN~2025wny.

\begin{acknowledgements}

    Each author group has authors ordered alphabetically and reflects equal contributions within the author group.
    AM acknowledges funding from the Deutsche Forschungsgemeinschaft (DFG, German Research Foundation) – SFB 1258 – 283604770.
    SS has received funding from the European Union’s Horizon 2022 research and innovation programme under the Marie Skłodowska-Curie grant agreement No 101105167 — FASTIDIoUS. We acknowledge financial support through grant PRIN-MIUR 2020SKSTHZ and from the University of Milan for the Nordic Optical Telescope programmes P68-804 and P71-804.
    AG, SHS and EM thank the Max Planck Society for support through the Max Planck Fellowship for SHS. 
    This work is supported in part by the Deutsche Forschungsgemeinschaft (DFG, German Research Foundation) under Germany's Excellence Strategy -- EXC-2094 -- 390783311. 
    MM acknowledges support by the SNSF (Swiss National Science Foundation) through return CH grant P5R5PT\_225598 and Ambizione grant PZ00P2\_223738.  
    TMR is part of the Cosmic Dawn Center (DAWN), which is funded by the Danish National Research Foundation under grant DNRF140. TMR and SM acknowledge support from the Research Council of Finland project 350458.
    TLK acknowledges support via a Warwick Astrophysics prize post-doctoral fellowship made possible thanks to a generous philanthropic donation.
    T.-W.C., AA, Y.-H.L. acknowledge the financial support from the Yushan Fellow Program by the Ministry of Education, Taiwan (MOE-111-YSFMS-0008-001-P1) and the National Science and Technology Council, Taiwan (NSTC grant 114-2112-M-008-021-MY3). 
    S.Y. acknowledges the funding from the National Natural Science Foundation of China under grant No. 12303046, the Startup Research Fund of Henan Academy of Sciences No. 242041217, and the Joint Fund of Henan Province Science and Technology R\&D Program No. 235200810057.
    YS acknowledges the support from the China Manned Space Program with grant no. CMS-CSST-2025-A20 and the National Natural Science Foundation of China (Grant No. 12333001).
    The Observatory of Maidanak team gratefully acknowledges the financial support provided by the Ministry of Higher Education, Science and Innovation of the Republic of Uzbekistan under grant No. IL-5421101855.
    GA wishes to acknowledge support from the United States Department of Energy, Office of Science, under Contract No.~DE-AC02-05CH11231.
    CPG acknowledges financial support from the Secretary of Universities and Research (Government of Catalonia) and by the Horizon 2020 Research and Innovation Programme of the European Union under the Marie Sk\l{}odowska-Curie and the Beatriu de Pin\'os 2021 BP 00168 programme, from the Spanish Ministerio de Ciencia e Innovaci\'on (MCIN) and the Agencia Estatal de Investigaci\'on (AEI) 10.13039/501100011033 under the PID2023-151307NB-I00 SNNEXT project, from Centro Superior de Investigaciones Cient\'ificas (CSIC) under the PIE project 20215AT016 and the program Unidad de Excelencia Mar\'ia de Maeztu CEX2020-001058-M, and from the Departament de Recerca i Universitats de la Generalitat de Catalunya through the 2021-SGR-01270 grant.
    This work was supported by the “Action Thématique de Physique Stellaire” (ATPS) of CNRS/INSU PN Astro cofunded by CEA and CNES.
    BS acknowledges the support of the French Agence Nationale de la Recherche (ANR), under grant ANR-23-CE31-0011 (project PEGaSUS).
    JDRP is supported by NASA through a Einstein Fellowship grant No. HF2-51541.001 awarded by the Space Telescope Science Institute (STScI), which is operated by the Association of Universities for Research in Astronomy, Inc., for NASA, under contract NAS5-26555.
    JJ acknowledges the support from National Natural Science Foundation of China (Grant No.~12393811), National Key R\&D Program of China (Grant No.~2023YFA1608100), the Strategic Priority Research Program of the Chinese Academy of Science (Grant No.~XDB0550300), and the Japan Society for the Promotion of Science (JSPS) KAKENHI grants JP22K14069.
    This research was partly based on observations made with the Nordic Optical Telescope (programme IDs: P72-503, P68-804, and P71-804) owned in collaboration by the University of Turku and Aarhus University, and operated jointly by Aarhus University, the University of Turku and the University of Oslo, representing Denmark, Finland and Norway, the University of Iceland and Stockholm University at the Observatorio del Roque de los Muchachos, La Palma, Spain, of the Instituto de Astrofisica de Canarias.
    The data presented here were obtained in part with ALFOSC, which is provided by the Instituto de Astrofisica de Andalucia (IAA) under a joint agreement with the University of Copenhagen and NOT.
    This publication makes use of data collected at the Lulin Observatory, which is partly supported by the TAOvA program under the NSTC grant 114-2740-M-008-002.

\end{acknowledgements}

\bibliographystyle{aa}
\bibliography{2025wny}

\begin{thebibliography}{86}
\expandafter\ifx\csname natexlab\endcsname\relax\def\natexlab#1{#1}\fi

\bibitem[{{Abbott} {et~al.}(2017){Abbott}, {Abbott}, {Abbott}, {Acernese}, {Ackley}, {Adams}, {Adams}, {Addesso}, {Adhikari}, {Adya}, {Affeldt}, {Afrough}, {Agarwal}, {Agathos}, {Agatsuma}, {Aggarwal}, {Aguiar}, {Aiello}, {Ain}, {Ajith}, {Allen}, {Allen}, {Allocca}, {Altin}, {Amato}, {Ananyeva}, {Anderson}, {Anderson}, {Angelova}, {Antier}, {Appert}, {Arai}, {Araya}, {Areeda}, {Arnaud}, {Arun}, {Ascenzi}, {Ashton}, {Ast}, {Aston}, {Astone}, {Atallah}, {Aufmuth}, {Aulbert}, {AultONeal}, {Austin}, {Avila-Alvarez}, {Babak}, {Bacon}, {Bader}, {Bae}, {Baker}, {Baldaccini}, {Ballardin}, {Ballmer}, {Banagiri}, {Barayoga}, {Barclay}, {Barish}, {Barker}, {Barkett}, {Barone}, {Barr}, {Barsotti}, {Barsuglia}, {Barta}, {Barthelmy}, {Bartlett}, {Bartos}, {Bassiri}, {Basti}, {Batch}, {Bawaj}, {Bayley}, {Bazzan}, {B{\'e}csy}, {Beer}, {Bejger}, {Belahcene}, {Bell}, {Berger}, {Bergmann}, {Bero}, {Berry}, {Bersanetti}, {Bertolini}, {Betzwieser}, {Bhagwat}, {Bhandare}, {Bilenko}, {Billingsley}, {Billman}, {Birch}, {Birney},
  {Birnholtz}, {Biscans}, {Biscoveanu}, {Bisht}, {Bitossi}, {Biwer}, {Bizouard}, {Blackburn}, {Blackman}, {Blair}, {Blair}, {Blair}, {Bloemen}, {Bock}, {Bode}, {Boer}, {Bogaert}, {Bohe}, {Bondu}, {Bonilla}, {Bonnand}, {Boom}, {Bork}, {Boschi}, {Bose}, {Bossie}, {Bouffanais}, {Bozzi}, {Bradaschia}, {Brady}, {Branchesi}, {Brau}, {Briant}, {Brillet}, {Brinkmann}, {Brisson}, {Brockill}, {Broida}, {Brooks}, {Brown}, {Brown}, {Brunett}, {Buchanan}, {Buikema}, {Bulik}, {Bulten}, {Buonanno}, {Buskulic}, {Buy}, {Byer}, {Cabero}, {Cadonati}, {Cagnoli}, {Cahillane}, {Calder{\'o}n Bustillo}, {Callister}, {Calloni}, {Camp}, {Canepa}, {Canizares}, {Cannon}, {Cao}, {Cao}, {Capano}, {Capocasa}, {Carbognani}, {Caride}, {Carney}, {Casanueva Diaz}, {Casentini}, {Caudill}, {Cavagli{\`a}}, {Cavalier}, {Cavalieri}, {Cella}, {Cepeda}, {Cerd{\'a}-Dur{\'a}n}, {Cerretani}, {Cesarini}, {Chamberlin}, {Chan}, {Chao}, {Charlton}, {Chase}, {Chassande-Mottin}, {Chatterjee}, {Chatziioannou}, {Cheeseboro}, {Chen}, {Chen}, {Chen}, {Cheng},
  {Chia}, {Chincarini}, {Chiummo}, {Chmiel}, {Cho}, {Cho}, {Chow}, {Christensen}, {Chu}, {Chua}, {Chua}, {Chung}, {Chung}, \& {Ciani}}]{abbott2017}
{Abbott}, B.~P., {Abbott}, R., {Abbott}, T.~D., {et~al.} 2017, \apjl, 848, L12

\bibitem[{{Adelman-McCarthy} {et~al.}(2007){Adelman-McCarthy}, {Ag{\"u}eros}, {Allam}, {Anderson}, {Anderson}, {Annis}, {Bahcall}, {Bailer-Jones}, {Baldry}, {Barentine}, {Beers}, {Belokurov}, {Berlind}, {Bernardi}, {Blanton}, {Bochanski}, {Boroski}, {Bramich}, {Brewington}, {Brinchmann}, {Brinkmann}, {Brunner}, {Budav{\'a}ri}, {Carey}, {Carliles}, {Carr}, {Castander}, {Connolly}, {Cool}, {Cunha}, {Csabai}, {Dalcanton}, {Doi}, {Eisenstein}, {Evans}, {Evans}, {Fan}, {Finkbeiner}, {Friedman}, {Frieman}, {Fukugita}, {Gillespie}, {Gilmore}, {Glazebrook}, {Gray}, {Grebel}, {Gunn}, {de Haas}, {Hall}, {Harvanek}, {Hawley}, {Hayes}, {Heckman}, {Hendry}, {Hennessy}, {Hindsley}, {Hirata}, {Hogan}, {Hogg}, {Holtzman}, {Ichikawa}, {Ichikawa}, {Ivezi{\'c}}, {Jester}, {Johnston}, {Jorgensen}, {Juri{\'c}}, {Kauffmann}, {Kent}, {Kleinman}, {Knapp}, {Kniazev}, {Kron}, {Krzesinski}, {Kuropatkin}, {Lamb}, {Lampeitl}, {Lee}, {Leger}, {Lima}, {Lin}, {Long}, {Loveday}, {Lupton}, {Mandelbaum}, {Margon}, {Mart{\'\i}nez-Delgado},
  {Matsubara}, {McGehee}, {McKay}, {Meiksin}, {Munn}, {Nakajima}, {Nash}, {Neilsen}, {Newberg}, {Nichol}, {Nieto-Santisteban}, {Nitta}, {Oyaizu}, {Okamura}, {Ostriker}, {Padmanabhan}, {Park}, {Peoples}, {Pier}, {Pope}, {Pourbaix}, {Quinn}, {Raddick}, {Re Fiorentin}, {Richards}, {Richmond}, {Rix}, {Rockosi}, {Schlegel}, {Schneider}, {Scranton}, {Seljak}, {Sheldon}, {Shimasaku}, {Silvestri}, {Smith}, {Smol{\v{c}}i{\'c}}, {Snedden}, {Stebbins}, {Stoughton}, {Strauss}, {SubbaRao}, {Suto}, {Szalay}, {Szapudi}, {Szkody}, {Tegmark}, {Thakar}, {Tremonti}, {Tucker}, {Uomoto}, {Vanden Berk}, {Vandenberg}, {Vidrih}, {Vogeley}, {Voges}, {Vogt}, {Weinberg}, {West}, {White}, {Wilhite}, {Yanny}, {Yocum}, {York}, {Zehavi}, {Zibetti}, \& {Zucker}}]{SDSS5}
{Adelman-McCarthy}, J.~K., {Ag{\"u}eros}, M.~A., {Allam}, S.~S., {et~al.} 2007, \apjs, 172, 634

\bibitem[{{Agrawal} {et~al.}(2025){Agrawal}, {Pierel}, {Narayan}, {Frye}, {Diego}, {Garuda}, {Grayling}, {Koekemoer}, {Mandel}, {Pascale}, {Vizgan}, \& {Windhorst}}]{agrawal25}
{Agrawal}, A., {Pierel}, J.~D.~R., {Narayan}, G., {et~al.} 2025, arXiv e-prints, arXiv:2510.07637

\bibitem[{{Arendse} {et~al.}(2024){Arendse}, {Dhawan}, {Sagu{\'e}s Carracedo}, {Peiris}, {Goobar}, {Wojtak}, {Alves}, {Biswas}, {Huber}, {Birrer}, \& {The LSST Dark Energy Science Collaboration}}]{Arendse+24}
{Arendse}, N., {Dhawan}, S., {Sagu{\'e}s Carracedo}, A., {et~al.} 2024, \mnras, 531, 3509

\bibitem[{{Bag} {et~al.}(2024){Bag}, {Huber}, {Suyu}, {Arendse}, {Andika}, {Ca{\~n}ameras}, {Kim}, {Linder}, {Lodha}, {Melo}, {More}, {Schuldt}, \& {Shafieloo}}]{Bag+24}
{Bag}, S., {Huber}, S., {Suyu}, S.~H., {et~al.} 2024, \aap, 691, A100

\bibitem[{{Basa} {et~al.}(2022){Basa}, {Lee}, {Dolon}, {Watson}, {Floriot}, {Atteia}, {Butler}, {Dornic}, {Lombardo}, {Ronayette}, {Ageron}, {Agneray}, {{\'A}ngeles}, {Bautista}, {Benamar-Aissa}, {Blanpain}, {Boulade}, {Boy}, {Buat}, {Cadena}, {Cuevas}, {Farah}, {Figueroa}, {Fuentes}, {Ga{\"\i}ti}, {Gallais}, {Kajfasz}, {Langarica}, {Langlois}, {Larrieu}, {Le Van Suu}, {Lecubin}, {L{\'o}pez {\'A}ngeles}, {Lugo}, {Malgoyre}, {Mathon}, {Moreau}, {Nouvel-De-La-Fl{\`e}che}, {Ochoa}, {Pedrayes-L{\'o}pez}, {Ramon}, {Ru{\'\i}z-D{\'\i}az-Soto}, {Tinoco}, \& {Valentin}}]{basa22}
{Basa}, S., {Lee}, W.~H., {Dolon}, F., {et~al.} 2022, in Society of Photo-Optical Instrumentation Engineers (SPIE) Conference Series, Vol. 12182, Ground-based and Airborne Telescopes IX, ed. H.~K. {Marshall}, J.~{Spyromilio}, \& T.~{Usuda}, 121821S

\bibitem[{{Bayer} {et~al.}(2021){Bayer}, {Huber}, {Vogl}, {Suyu}, {Taubenberger}, {Sluse}, {Chan}, \& {Kerzendorf}}]{Bayer21}
{Bayer}, J., {Huber}, S., {Vogl}, C., {et~al.} 2021, \aap, 653, A29

\bibitem[{{Birrer} {et~al.}(2020){Birrer}, {Shajib}, {Galan}, {Millon}, {Treu}, {Agnello}, {Auger}, {Chen}, {Christensen}, {Collett}, {Courbin}, {Fassnacht}, {Koopmans}, {Marshall}, {Park}, {Rusu}, {Sluse}, {Spiniello}, {Suyu}, {Wagner-Carena}, {Wong}, {Barnab{\`e}}, {Bolton}, {Czoske}, {Ding}, {Frieman}, \& {Van de Vyvere}}]{Birrer20}
{Birrer}, S., {Shajib}, A.~J., {Galan}, A., {et~al.} 2020, \aap, 643, A165

\bibitem[{{Blondin} \& {Tonry}(2007)}]{blondin2007}
{Blondin}, S. \& {Tonry}, J.~L. 2007, \apj, 666, 1024

\bibitem[{{Bongard} {et~al.}(2011){Bongard}, {Soulez}, {Thi{\'e}baut}, \& {Pecontal}}]{bongard2011}
{Bongard}, S., {Soulez}, F., {Thi{\'e}baut}, {\'E}., \& {Pecontal}, {\'E}. 2011, \mnras, 418, 258

\bibitem[{{Brennan} \& {Fraser}(2022)}]{2022A&A...667A..62B}
{Brennan}, S.~J. \& {Fraser}, M. 2022, \aap, 667, A62

\bibitem[{{Buton} {et~al.}(2013){Buton}, {Copin}, {Aldering}, {Antilogus}, {Aragon}, {Bailey}, {Baltay}, {Bongard}, {Canto}, {Cellier-Holzem}, {Childress}, {Chotard}, {Fakhouri}, {Gangler}, {Guy}, {Hsiao}, {Kerschhaggl}, {Kowalski}, {Loken}, {Nugent}, {Paech}, {Pain}, {P{\'e}contal}, {Pereira}, {Perlmutter}, {Rabinowitz}, {Rigault}, {Runge}, {Scalzo}, {Smadja}, {Tao}, {Thomas}, {Weaver}, {Wu}, \& {Nearby SuperNova Factory}}]{buton2013}
{Buton}, C., {Copin}, Y., {Aldering}, G., {et~al.} 2013, \aap, 549, A8

\bibitem[{{Ca{\~n}ameras} {et~al.}(2020){Ca{\~n}ameras}, {Schuldt}, {Suyu}, {Taubenberger}, {Meinhardt}, {Leal-Taix{\'e}}, {Lemon}, {Rojas}, \& {Savary}}]{canameras20}
{Ca{\~n}ameras}, R., {Schuldt}, S., {Suyu}, S.~H., {et~al.} 2020, \aap, 644, A163

\bibitem[{{Chambers} {et~al.}(2016){Chambers}, {Magnier}, {Metcalfe}, {Flewelling}, {Huber}, {Waters}, {Denneau}, {Draper}, {Farrow}, {Finkbeiner}, {Holmberg}, {Koppenhoefer}, {Price}, {Rest}, {Saglia}, {Schlafly}, {Smartt}, {Sweeney}, {Wainscoat}, {Burgett}, {Chastel}, {Grav}, {Heasley}, {Hodapp}, {Jedicke}, {Kaiser}, {Kudritzki}, {Luppino}, {Lupton}, {Monet}, {Morgan}, {Onaka}, {Shiao}, {Stubbs}, {Tonry}, {White}, {Ba{\~n}ados}, {Bell}, {Bender}, {Bernard}, {Boegner}, {Boffi}, {Botticella}, {Calamida}, {Casertano}, {Chen}, {Chen}, {Cole}, {Deacon}, {Frenk}, {Fitzsimmons}, {Gezari}, {Gibbs}, {Goessl}, {Goggia}, {Gourgue}, {Goldman}, {Grant}, {Grebel}, {Hambly}, {Hasinger}, {Heavens}, {Heckman}, {Henderson}, {Henning}, {Holman}, {Hopp}, {Ip}, {Isani}, {Jackson}, {Keyes}, {Koekemoer}, {Kotak}, {Le}, {Liska}, {Long}, {Lucey}, {Liu}, {Martin}, {Masci}, {McLean}, {Mindel}, {Misra}, {Morganson}, {Murphy}, {Obaika}, {Narayan}, {Nieto-Santisteban}, {Norberg}, {Peacock}, {Pier}, {Postman}, {Primak}, {Rae}, {Rai},
  {Riess}, {Riffeser}, {Rix}, {R{\"o}ser}, {Russel}, {Rutz}, {Schilbach}, {Schultz}, {Scolnic}, {Strolger}, {Szalay}, {Seitz}, {Small}, {Smith}, {Soderblom}, {Taylor}, {Thomson}, {Taylor}, {Thakar}, {Thiel}, {Thilker}, {Unger}, {Urata}, {Valenti}, {Wagner}, {Walder}, {Walter}, {Watters}, {Werner}, {Wood-Vasey}, \& {Wyse}}]{2016chambers}
{Chambers}, K.~C., {Magnier}, E.~A., {Metcalfe}, N., {et~al.} 2016, arXiv e-prints, arXiv:1612.05560

\bibitem[{{Chen} {et~al.}(2022){Chen}, {Hashimoto}, {Goto}, {Kim}, {Santos}, {On}, {Lu}, \& {Hsiao}}]{chen22}
{Chen}, B.~H., {Hashimoto}, T., {Goto}, T., {et~al.} 2022, \mnras, 509, 1227

\bibitem[{{Chen} {et~al.}(2013){Chen}, {Smartt}, {Bresolin}, {Pastorello}, {Kudritzki}, {Kotak}, {McCrum}, {Fraser}, \& {Valenti}}]{2013ApJ...763L..28C}
{Chen}, T.-W., {Smartt}, S.~J., {Bresolin}, F., {et~al.} 2013, \apjl, 763, L28

\bibitem[{{Cooke} {et~al.}(2012){Cooke}, {Sullivan}, {Gal-Yam}, {Barton}, {Carlberg}, {Ryan-Weber}, {Horst}, {Omori}, \& {D{\'\i}az}}]{cooke2012}
{Cooke}, J., {Sullivan}, M., {Gal-Yam}, A., {et~al.} 2012, \nat, 491, 228

\bibitem[{{Curtin} {et~al.}(2019){Curtin}, {Cooke}, {Moriya}, {Tanaka}, {Quimby}, {Bernard}, {Galbany}, {Jiang}, {Lee}, {Maeda}, {Morokuma}, {Nomoto}, {Pignata}, {Pritchard}, {Suzuki}, {Takahashi}, {Tanaka}, {Tominaga}, {Yamaguchi}, \& {Yasuda}}]{curtin2019}
{Curtin}, C., {Cooke}, J., {Moriya}, T.~J., {et~al.} 2019, \apjs, 241, 17

\bibitem[{{DESI Collaboration} {et~al.}(2016){DESI Collaboration}, {Aghamousa}, {Aguilar}, {Ahlen}, {Alam}, {Allen}, {Allende Prieto}, {Annis}, {Bailey}, {Balland}, {Ballester}, {Baltay}, {Beaufore}, {Bebek}, {Beers}, {Bell}, {Bernal}, {Besuner}, {Beutler}, {Blake}, {Bleuler}, {Blomqvist}, {Blum}, {Bolton}, {Briceno}, {Brooks}, {Brownstein}, {Buckley-Geer}, {Burden}, {Burtin}, {Busca}, {Cahn}, {Cai}, {Cardiel-Sas}, {Carlberg}, {Carton}, {Casas}, {Castander}, {Cervantes-Cota}, {Claybaugh}, {Close}, {Coker}, {Cole}, {Comparat}, {Cooper}, {Cousinou}, {Crocce}, {Cuby}, {Cunningham}, {Davis}, {Dawson}, {de la Macorra}, {De Vicente}, {Delubac}, {Derwent}, {Dey}, {Dhungana}, {Ding}, {Doel}, {Duan}, {Ealet}, {Edelstein}, {Eftekharzadeh}, {Eisenstein}, {Elliott}, {Escoffier}, {Evatt}, {Fagrelius}, {Fan}, {Fanning}, {Farahi}, {Farihi}, {Favole}, {Feng}, {Fernandez}, {Findlay}, {Finkbeiner}, {Fitzpatrick}, {Flaugher}, {Flender}, {Font-Ribera}, {Forero-Romero}, {Fosalba}, {Frenk}, {Fumagalli}, {Gaensicke}, {Gallo},
  {Garcia-Bellido}, {Gaztanaga}, {Pietro Gentile Fusillo}, {Gerard}, {Gershkovich}, {Giannantonio}, {Gillet}, {Gonzalez-de-Rivera}, {Gonzalez-Perez}, {Gott}, {Graur}, {Gutierrez}, {Guy}, {Habib}, {Heetderks}, {Heetderks}, {Heitmann}, {Hellwing}, {Herrera}, {Ho}, {Holland}, {Honscheid}, {Huff}, {Hutchinson}, {Huterer}, {Hwang}, {Illa Laguna}, {Ishikawa}, {Jacobs}, {Jeffrey}, {Jelinsky}, {Jennings}, {Jiang}, {Jimenez}, {Johnson}, {Joyce}, {Jullo}, {Juneau}, {Kama}, {Karcher}, {Karkar}, {Kehoe}, {Kennamer}, {Kent}, {Kilbinger}, {Kim}, {Kirkby}, {Kisner}, {Kitanidis}, {Kneib}, {Koposov}, {Kovacs}, {Koyama}, {Kremin}, {Kron}, {Kronig}, {Kueter-Young}, {Lacey}, {Lafever}, {Lahav}, {Lambert}, {Lampton}, {Landriau}, {Lang}, {Lauer}, {Le Goff}, {Le Guillou}, {Le Van Suu}, {Lee}, {Lee}, {Leitner}, {Lesser}, {Levi}, {L'Huillier}, {Li}, {Liang}, {Lin}, {Linder}, {Loebman}, {Luki{\'c}}, {Ma}, {MacCrann}, {Magneville}, {Makarem}, {Manera}, {Manser}, {Marshall}, {Martini}, {Massey}, {Matheson}, {McCauley}, {McDonald},
  {McGreer}, {Meisner}, {Metcalfe}, {Miller}, {Miquel}, {Moustakas}, {Myers}, {Naik}, {Newman}, {Nichol}, {Nicola}, {Nicolati da Costa}, {Nie}, {Niz}, {Norberg}, {Nord}, {Norman}, {Nugent}, {O'Brien}, {Oh}, \& {Olsen}}]{2016DESIcollab}
{DESI Collaboration}, {Aghamousa}, A., {Aguilar}, J., {et~al.} 2016, arXiv e-prints, arXiv:1611.00036

\bibitem[{{Dey} {et~al.}(2019){Dey}, {Schlegel}, {Lang}, {Blum}, {Burleigh}, {Fan}, {Findlay}, {Finkbeiner}, {Herrera}, {Juneau}, {Landriau}, {Levi}, {McGreer}, {Meisner}, {Myers}, {Moustakas}, {Nugent}, {Patej}, {Schlafly}, {Walker}, {Valdes}, {Weaver}, {Y{\`e}che}, {Zou}, {Zhou}, {Abareshi}, {Abbott}, {Abolfathi}, {Aguilera}, {Alam}, {Allen}, {Alvarez}, {Annis}, {Ansarinejad}, {Aubert}, {Beechert}, {Bell}, {BenZvi}, {Beutler}, {Bielby}, {Bolton}, {Brice{\~n}o}, {Buckley-Geer}, {Butler}, {Calamida}, {Carlberg}, {Carter}, {Casas}, {Castander}, {Choi}, {Comparat}, {Cukanovaite}, {Delubac}, {DeVries}, {Dey}, {Dhungana}, {Dickinson}, {Ding}, {Donaldson}, {Duan}, {Duckworth}, {Eftekharzadeh}, {Eisenstein}, {Etourneau}, {Fagrelius}, {Farihi}, {Fitzpatrick}, {Font-Ribera}, {Fulmer}, {G{\"a}nsicke}, {Gaztanaga}, {George}, {Gerdes}, {Gontcho}, {Gorgoni}, {Green}, {Guy}, {Harmer}, {Hernandez}, {Honscheid}, {Huang}, {James}, {Jannuzi}, {Jiang}, {Joyce}, {Karcher}, {Karkar}, {Kehoe}, {Kneib}, {Kueter-Young}, {Lan},
  {Lauer}, {Le Guillou}, {Le Van Suu}, {Lee}, {Lesser}, {Perreault Levasseur}, {Li}, {Mann}, {Marshall}, {Mart{\'\i}nez-V{\'a}zquez}, {Martini}, {du Mas des Bourboux}, {McManus}, {Meier}, {M{\'e}nard}, {Metcalfe}, {Mu{\~n}oz-Guti{\'e}rrez}, {Najita}, {Napier}, {Narayan}, {Newman}, {Nie}, {Nord}, {Norman}, {Olsen}, {Paat}, {Palanque-Delabrouille}, {Peng}, {Poppett}, {Poremba}, {Prakash}, {Rabinowitz}, {Raichoor}, {Rezaie}, {Robertson}, {Roe}, {Ross}, {Ross}, {Rudnick}, {Safonova}, {Saha}, {S{\'a}nchez}, {Savary}, {Schweiker}, {Scott}, {Seo}, {Shan}, {Silva}, {Slepian}, {Soto}, {Sprayberry}, {Staten}, {Stillman}, {Stupak}, {Summers}, {Sien Tie}, {Tirado}, {Vargas-Maga{\~n}a}, {Vivas}, {Wechsler}, {Williams}, {Yang}, {Yang}, {Yapici}, {Zaritsky}, {Zenteno}, {Zhang}, {Zhang}, {Zhou}, \& {Zhou}}]{2019Dey}
{Dey}, A., {Schlegel}, D.~J., {Lang}, D., {et~al.} 2019, \aj, 157, 168

\bibitem[{{Dhawan} {et~al.}(2020){Dhawan}, {Johansson}, {Goobar}, {Amanullah}, {M{\"o}rtsell}, {Cenko}, {Cooray}, {Fox}, {Goldstein}, {Kalender}, {Kasliwal}, {Kulkarni}, {Lee}, {Nayyeri}, {Nugent}, {Ofek}, \& {Quimby}}]{dhawan20}
{Dhawan}, S., {Johansson}, J., {Goobar}, A., {et~al.} 2020, \mnras, 491, 2639

\bibitem[{{Dhawan} {et~al.}(2024){Dhawan}, {Pierel}, {Gu}, {Newman}, {Larison}, {Siebert}, {Petrushevska}, {Poidevin}, {Jha}, {Chen}, {Ellis}, {Frye}, {Hjorth}, {Koekemoer}, {P{\'e}rez-Fournon}, {Rest}, {Treu}, {Windhorst}, \& {Zenati}}]{Dhawan+24}
{Dhawan}, S., {Pierel}, J.~D.~R., {Gu}, M., {et~al.} 2024, \mnras, 535, 2939

\bibitem[{{Ehgamberdiev}(2018)}]{2018NatAs}
{Ehgamberdiev}, S. 2018, Nature Astronomy, 2, 349

\bibitem[{{Frohmaier} {et~al.}(2021){Frohmaier}, {Angus}, {Vincenzi}, {Sullivan}, {Smith}, {Nugent}, {Cenko}, {Gal-Yam}, {Kulkarni}, {Law}, \& {Quimby}}]{2021MNRAS.500.5142F}
{Frohmaier}, C., {Angus}, C.~R., {Vincenzi}, M., {et~al.} 2021, \mnras, 500, 5142

\bibitem[{{Frye} {et~al.}(2023){Frye}, {Pascale}, {Cohen}, {Summers}, {Foo}, {Kamieneski}, {Carleton}, {Jansen}, {Pierel}, {Engesser}, {Chen}, {Austin}, {Marshall}, {Trussler}, {Meena}, {Leimbach}, {Garuda}, {Honor}, {Furtak}, {Strolger}, {Windhorst}, {Koekemoer}, {Zitrin}, {Diego}, {Kelly}, {Coe}, {Conselice}, {Dai}, {D{\^a}Silva}, {Dole}, {Driver}, {Grogin}, {Nonino}, {Pirzkal}, {Polletta}, {Robotham}, {Rutkowski}, {Ryan}, {Tompkins}, {Willmer}, {Willner}, {Yan}, \& {Yun}}]{2023Frye}
{Frye}, B., {Pascale}, M., {Cohen}, S., {et~al.} 2023, Transient Name Server AstroNote, 96, 1

\bibitem[{{Gaia Collaboration} {et~al.}(2021){Gaia Collaboration}, {Brown}, {Vallenari}, {Prusti}, {de Bruijne}, {Babusiaux}, {Biermann}, {Creevey}, {Evans}, {Eyer}, {Hutton}, {Jansen}, {Jordi}, {Klioner}, {Lammers}, {Lindegren}, {Luri}, {Mignard}, {Panem}, {Pourbaix}, {Randich}, {Sartoretti}, {Soubiran}, {Walton}, {Arenou}, {Bailer-Jones}, {Bastian}, {Cropper}, {Drimmel}, {Katz}, {Lattanzi}, {van Leeuwen}, {Bakker}, {Cacciari}, {Casta{\~n}eda}, {De Angeli}, {Ducourant}, {Fabricius}, {Fouesneau}, {Fr{\'e}mat}, {Guerra}, {Guerrier}, {Guiraud}, {Jean-Antoine Piccolo}, {Masana}, {Messineo}, {Mowlavi}, {Nicolas}, {Nienartowicz}, {Pailler}, {Panuzzo}, {Riclet}, {Roux}, {Seabroke}, {Sordo}, {Tanga}, {Th{\'e}venin}, {Gracia-Abril}, {Portell}, {Teyssier}, {Altmann}, {Andrae}, {Bellas-Velidis}, {Benson}, {Berthier}, {Blomme}, {Brugaletta}, {Burgess}, {Busso}, {Carry}, {Cellino}, {Cheek}, {Clementini}, {Damerdji}, {Davidson}, {Delchambre}, {Dell'Oro}, {Fern{\'a}ndez-Hern{\'a}ndez}, {Galluccio}, {Garc{\'\i}a-Lario},
  {Garcia-Reinaldos}, {Gonz{\'a}lez-N{\'u}{\~n}ez}, {Gosset}, {Haigron}, {Halbwachs}, {Hambly}, {Harrison}, {Hatzidimitriou}, {Heiter}, {Hern{\'a}ndez}, {Hestroffer}, {Hodgkin}, {Holl}, {Jan{\ss}en}, {Jevardat de Fombelle}, {Jordan}, {Krone-Martins}, {Lanzafame}, {L{\"o}ffler}, {Lorca}, {Manteiga}, {Marchal}, {Marrese}, {Moitinho}, {Mora}, {Muinonen}, {Osborne}, {Pancino}, {Pauwels}, {Petit}, {Recio-Blanco}, {Richards}, {Riello}, {Rimoldini}, {Robin}, {Roegiers}, {Rybizki}, {Sarro}, {Siopis}, {Smith}, {Sozzetti}, {Ulla}, {Utrilla}, {van Leeuwen}, {van Reeven}, {Abbas}, {Abreu Aramburu}, {Accart}, {Aerts}, {Aguado}, {Ajaj}, {Altavilla}, {{\'A}lvarez}, {{\'A}lvarez Cid-Fuentes}, {Alves}, {Anderson}, {Anglada Varela}, {Antoja}, {Audard}, {Baines}, {Baker}, {Balaguer-N{\'u}{\~n}ez}, {Balbinot}, {Balog}, {Barache}, {Barbato}, {Barros}, {Barstow}, {Bartolom{\'e}}, {Bassilana}, {Bauchet}, {Baudesson-Stella}, {Becciani}, {Bellazzini}, {Bernet}, {Bertone}, {Bianchi}, {Blanco-Cuaresma}, {Boch}, {Bombrun}, {Bossini},
  {Bouquillon}, {Bragaglia}, {Bramante}, {Breedt}, {Bressan}, {Brouillet}, {Bucciarelli}, {Burlacu}, {Busonero}, {Butkevich}, {Buzzi}, {Caffau}, {Cancelliere}, {C{\'a}novas}, {Cantat-Gaudin}, {Carballo}, {Carlucci}, {Carnerero}, {Carrasco}, {Casamiquela}, {Castellani}, {Castro-Ginard}, {Castro Sampol}, {Chaoul}, {Charlot}, {Chemin}, {Chiavassa}, {Cioni}, {Comoretto}, {Cooper}, {Cornez}, {Cowell}, {Crifo}, {Crosta}, {Crowley}, {Dafonte}, {Dapergolas}, {David}, \& {David}}]{2021A&A...649A...1G}
{Gaia Collaboration}, {Brown}, A.~G.~A., {Vallenari}, A., {et~al.} 2021, \aap, 649, A1

\bibitem[{{Goldwasser} {et~al.}(2022){Goldwasser}, {Yaron}, {Sass}, {Irani}, {Gal-Yam}, \& {Howell}}]{goldwasser2022}
{Goldwasser}, S., {Yaron}, O., {Sass}, A., {et~al.} 2022, Transient Name Server AstroNote, 191, 1

\bibitem[{{Gomez} {et~al.}(2024){Gomez}, {Nicholl}, {Berger}, {Blanchard}, {Villar}, {Rest}, {Hosseinzadeh}, {Aamer}, {Ajay}, {Athukoralalage}, {Coulter}, {Eftekhari}, {Fiore}, {Franz}, {Fox}, {Gagliano}, {Hiramatsu}, {Howell}, {Hsu}, {Karmen}, {Siebert}, {K{\"o}nyves-T{\'o}th}, {Kumar}, {McCully}, {Pellegrino}, {Pierel}, {Rest}, \& {Wang}}]{gomez2024}
{Gomez}, S., {Nicholl}, M., {Berger}, E., {et~al.} 2024, \mnras, 535, 471

\bibitem[{{Goobar} {et~al.}(2017){Goobar}, {Amanullah}, {Kulkarni}, {Nugent}, {Johansson}, {Steidel}, {Law}, {M{\"o}rtsell}, {Quimby}, {Blagorodnova}, {Brand eker}, {Cao}, {Cooray}, {Ferretti}, {Fremling}, {Hangard}, {Kasliwal}, {Kupfer}, {Lunnan}, {Masci}, {Miller}, {Nayyeri}, {Neill}, {Ofek}, {Papadogiannakis}, {Petrushevska}, {Ravi}, {Sollerman}, {Sullivan}, {Taddia}, {Walters}, {Wilson}, {Yan}, \& {Yaron}}]{goobar17}
{Goobar}, A., {Amanullah}, R., {Kulkarni}, S.~R., {et~al.} 2017, Science, 356, 291

\bibitem[{{Goobar} {et~al.}(2023){Goobar}, {Johansson}, {Schulze}, {Arendse}, {Carracedo}, {Dhawan}, {M{\"o}rtsell}, {Fremling}, {Yan}, {Perley}, {Sollerman}, {Joseph}, {Hinds}, {Meynardie}, {Andreoni}, {Bellm}, {Bloom}, {Collett}, {Drake}, {Graham}, {Kasliwal}, {Kulkarni}, {Lemon}, {Miller}, {Neill}, {Nordin}, {Pierel}, {Richard}, {Riddle}, {Rigault}, {Rusholme}, {Sharma}, {Stein}, {Stewart}, {Townsend}, {Vinko}, {Wheeler}, \& {Wold}}]{2023goobar}
{Goobar}, A., {Johansson}, J., {Schulze}, S., {et~al.} 2023, Nature Astronomy, 7, 1098

\bibitem[{{Grillo} {et~al.}(2024){Grillo}, {Pagano}, {Rosati}, \& {Suyu}}]{grillo24}
{Grillo}, C., {Pagano}, L., {Rosati}, P., \& {Suyu}, S.~H. 2024, \aap, 684, L23

\bibitem[{{Gwyn}(2012)}]{2012Gwyn}
{Gwyn}, S. D.~J. 2012, \aj, 143, 38

\bibitem[{{Harutyunyan} {et~al.}(2008){Harutyunyan}, {Pfahler}, {Pastorello}, {Taubenberger}, {Turatto}, {Cappellaro}, {Benetti}, {Elias-Rosa}, {Navasardyan}, {Valenti}, {Stanishev}, {Patat}, {Riello}, {Pignata}, \& {Hillebrandt}}]{harutyunyan2008}
{Harutyunyan}, A.~H., {Pfahler}, P., {Pastorello}, A., {et~al.} 2008, \aap, 488, 383

\bibitem[{{Hopp} {et~al.}(2014){Hopp}, {Bender}, {Grupp}, {Goessl}, {Lang-Bardl}, {Mitsch}, {Riffeser}, \& {Ageorges}}]{2014SPIE.9145E..2DH}
{Hopp}, U., {Bender}, R., {Grupp}, F., {et~al.} 2014, in Society of Photo-Optical Instrumentation Engineers (SPIE) Conference Series, Vol. 9145, Ground-based and Airborne Telescopes V, ed. L.~M. {Stepp}, R.~{Gilmozzi}, \& H.~J. {Hall}, 91452D

\bibitem[{{Howell} {et~al.}(2013){Howell}, {Kasen}, {Lidman}, {Sullivan}, {Conley}, {Astier}, {Balland}, {Carlberg}, {Fouchez}, {Guy}, {Hardin}, {Pain}, {Palanque-Delabrouille}, {Perrett}, {Pritchet}, {Regnault}, {Rich}, \& {Ruhlmann-Kleider}}]{howell2013}
{Howell}, D.~A., {Kasen}, D., {Lidman}, C., {et~al.} 2013, \apj, 779, 98

\bibitem[{{Howell} {et~al.}(2005){Howell}, {Sullivan}, {Perrett}, {Bronder}, {Hook}, {Astier}, {Aubourg}, {Balam}, {Basa}, {Carlberg}, {Fabbro}, {Fouchez}, {Guy}, {Lafoux}, {Neill}, {Pain}, {Palanque-Delabrouille}, {Pritchet}, {Regnault}, {Rich}, {Taillet}, {Knop}, {McMahon}, {Perlmutter}, \& {Walton}}]{howell2005}
{Howell}, D.~A., {Sullivan}, M., {Perrett}, K., {et~al.} 2005, \apj, 634, 1190

\bibitem[{{Howell}(1989)}]{howell1989}
{Howell}, S.~B. 1989, \pasp, 101, 616

\bibitem[{{Im} {et~al.}(2010){Im}, {Ko}, {Cho}, {Choi}, {Jeon}, {Lee}, \& {Ibrahimov}}]{2010JKAS}
{Im}, M.-S., {Ko}, J.-W., {Cho}, Y.-S., {et~al.} 2010, Journal of Korean Astronomical Society, 43, 75

\bibitem[{{Inserra} {et~al.}(2013){Inserra}, {Smartt}, {Jerkstrand}, {Valenti}, {Fraser}, {Wright}, {Smith}, {Chen}, {Kotak}, {Pastorello}, {Nicholl}, {Bresolin}, {Kudritzki}, {Benetti}, {Botticella}, {Burgett}, {Chambers}, {Ergon}, {Flewelling}, {Fynbo}, {Geier}, {Hodapp}, {Howell}, {Huber}, {Kaiser}, {Leloudas}, {Magill}, {Magnier}, {McCrum}, {Metcalfe}, {Price}, {Rest}, {Sollerman}, {Sweeney}, {Taddia}, {Taubenberger}, {Tonry}, {Wainscoat}, {Waters}, \& {Young}}]{inserra2013}
{Inserra}, C., {Smartt}, S.~J., {Jerkstrand}, A., {et~al.} 2013, \apj, 770, 128

\bibitem[{{Johansson} {et~al.}(2021){Johansson}, {Goobar}, {Price}, {Sagu{\'e}s Carracedo}, {Della Bruna}, {Nugent}, {Dhawan}, {M{\"o}rtsell}, {Papadogiannakis}, {Amanullah}, {Goldstein}, {Cenko}, {De}, {Dugas}, {Kasliwal}, {Kulkarni}, \& {Lunnan}}]{Johansson21}
{Johansson}, J., {Goobar}, A., {Price}, S.~H., {et~al.} 2021, \mnras, 502, 510

\bibitem[{{Johansson} {et~al.}(2025){Johansson}, {Qin}, {Goobar}, {Perley}, {Wise}, \& {Lemon}}]{johansson2025}
{Johansson}, J., {Qin}, Y.-J., {Goobar}, A., {et~al.} 2025, Transient Name Server AstroNote, 306, 1

\bibitem[{{Kasen} {et~al.}(2006){Kasen}, {Thomas}, \& {Nugent}}]{kasen2006}
{Kasen}, D., {Thomas}, R.~C., \& {Nugent}, P. 2006, \apj, 651, 366

\bibitem[{{Kelly} {et~al.}(2023){Kelly}, {Rodney}, {Treu}, {Oguri}, {Chen}, {Zitrin}, {Birrer}, {Bonvin}, {Dessart}, {Diego}, {Filippenko}, {Foley}, {Gilman}, {Hjorth}, {Jauzac}, {Mandel}, {Millon}, {Pierel}, {Sharon}, {Thorp}, {Williams}, {Broadhurst}, {Dressler}, {Graur}, {Jha}, {McCully}, {Postman}, {Schmidt}, {Tucker}, \& {von der Linden}}]{Kelly23}
{Kelly}, P.~L., {Rodney}, S., {Treu}, T., {et~al.} 2023, Science, 380, abh1322

\bibitem[{{Kelly} {et~al.}(2015){Kelly}, {Rodney}, {Treu}, {Foley}, {Brammer}, {Schmidt}, {Zitrin}, {Sonnenfeld}, {Strolger}, {Graur}, {Filippenko}, {Jha}, {Riess}, {Bradac}, {Weiner}, {Scolnic}, {Malkan}, {von der Linden}, {Trenti}, {Hjorth}, {Gavazzi}, {Fontana}, {Merten}, {McCully}, {Jones}, {Postman}, {Dressler}, {Patel}, {Cenko}, {Graham}, \& {Tucker}}]{Kelly15}
{Kelly}, P.~L., {Rodney}, S.~A., {Treu}, T., {et~al.} 2015, Science, 347, 1123

\bibitem[{{King} {et~al.}(2013){King}, {Naylor}, {Broos}, {Getman}, \& {Feigelson}}]{king2013}
{King}, R.~R., {Naylor}, T., {Broos}, P.~S., {Getman}, K.~V., \& {Feigelson}, E.~D. 2013, \apjs, 209, 28

\bibitem[{Lang-Bardl {et~al.}(2016)Lang-Bardl, Bender, Goessl, Grupp, Hess, Kaminski, Hodapp, Hopp, Jacobson, Kravcar, {et~al.}}]{lang2016wendelstein}
Lang-Bardl, F., Bender, R., Goessl, C., {et~al.} 2016, in Ground-based and Airborne Instrumentation for Astronomy VI, Vol. 9908, SPIE, 1295--1302

\bibitem[{{Lantz} {et~al.}(2004){Lantz}, {Aldering}, {Antilogus}, {Bonnaud}, {Capoani}, {Castera}, {Copin}, {Dubet}, {Gangler}, {Henault}, {Lemonnier}, {Pain}, {Pecontal}, {Pecontal}, \& {Smadja}}]{lantz2004}
{Lantz}, B., {Aldering}, G., {Antilogus}, P., {et~al.} 2004, in Society of Photo-Optical Instrumentation Engineers (SPIE) Conference Series, Vol. 5249, Optical Design and Engineering, ed. L.~{Mazuray}, P.~J. {Rogers}, \& R.~{Wartmann}, 146--155

\bibitem[{{Levi} {et~al.}(2013){Levi}, {Bebek}, {Beers}, {Blum}, {Cahn}, {Eisenstein}, {Flaugher}, {Honscheid}, {Kron}, {Lahav}, {McDonald}, {Roe}, {Schlegel}, \& {representing the DESI collaboration}}]{2013Levi}
{Levi}, M., {Bebek}, C., {Beers}, T., {et~al.} 2013, arXiv e-prints, arXiv:1308.0847

\bibitem[{{Liu} \& {Oguri}(2025)}]{LiuOguri25}
{Liu}, Y. \& {Oguri}, M. 2025, \prd, 111, 123506

\bibitem[{{Mazzali} {et~al.}(2016){Mazzali}, {Sullivan}, {Pian}, {Greiner}, \& {Kann}}]{mazzali2016}
{Mazzali}, P.~A., {Sullivan}, M., {Pian}, E., {Greiner}, J., \& {Kann}, D.~A. 2016, \mnras, 458, 3455

\bibitem[{{Messa} {et~al.}(2025){Messa}, {Vanzella}, {Loiacono}, {Bergamini}, {Castellano}, {Sun}, {Willott}, {Windhorst}, {Yan}, {Angora}, {Rosati}, {Adamo}, {Annibali}, {Bolamperti}, {Brada{\v{c}}}, {Bradley}, {Calura}, {Claeyssens}, {Comastri}, {Conselice}, {D'Silva}, {Dickinson}, {Frye}, {Grillo}, {Grogin}, {Gruppioni}, {Koekemoer}, {Meneghetti}, {Me{\v{s}}tri{\'c}}, {Pascale}, {Ravindranath}, {Ricotti}, {Summers}, \& {Zanella}}]{messa25}
{Messa}, M., {Vanzella}, E., {Loiacono}, F., {et~al.} 2025, \aap, 694, A59

\bibitem[{{Me{\v{s}}tri{\'c}} {et~al.}(2023){Me{\v{s}}tri{\'c}}, {Vanzella}, {Upadhyaya}, {Martins}, {Marques-Chaves}, {Schaerer}, {Guibert}, {Zanella}, {Grillo}, {Rosati}, {Calura}, {Caminha}, {Bolamperti}, {Meneghetti}, {Bergamini}, {Mercurio}, {Nonino}, \& {Pascale}}]{mestric23}
{Me{\v{s}}tri{\'c}}, U., {Vanzella}, E., {Upadhyaya}, A., {et~al.} 2023, \aap, 673, A50

\bibitem[{{Millon} {et~al.}(2020){Millon}, {Courbin}, {Bonvin}, {Buckley-Geer}, {Fassnacht}, {Frieman}, {Marshall}, {Suyu}, {Treu}, {Anguita}, {Motta}, {Agnello}, {Chan}, {Chao}, {Chijani}, {Gilman}, {Gilmore}, {Lemon}, {Lucey}, {Melo}, {Paic}, {Rojas}, {Sluse}, {Williams}, {Hempel}, {Kim}, {Lachaume}, \& {Rabus}}]{millon20a}
{Millon}, M., {Courbin}, F., {Bonvin}, V., {et~al.} 2020, \aap, 642, A193

\bibitem[{{Moffat}(1969)}]{1969A&A.....3..455M}
{Moffat}, A.~F.~J. 1969, \aap, 3, 455

\bibitem[{{Moresco} {et~al.}(2022){Moresco}, {Amati}, {Amendola}, {Birrer}, {Blakeslee}, {Cantiello}, {Cimatti}, {Darling}, {Della Valle}, {Fishbach}, {Grillo}, {Hamaus}, {Holz}, {Izzo}, {Jimenez}, {Lusso}, {Meneghetti}, {Piedipalumbo}, {Pisani}, {Pourtsidou}, {Pozzetti}, {Quartin}, {Risaliti}, {Rosati}, \& {Verde}}]{moresco22}
{Moresco}, M., {Amati}, L., {Amendola}, L., {et~al.} 2022, Living Reviews in Relativity, 25, 6

\bibitem[{{Oguri}(2019)}]{Oguri19}
{Oguri}, M. 2019, Reports on Progress in Physics, 82, 126901

\bibitem[{{Oguri} \& {Marshall}(2010)}]{oguri2010}
{Oguri}, M. \& {Marshall}, P.~J. 2010, \mnras, 405, 2579

\bibitem[{{Pascale} {et~al.}(2025){Pascale}, {Frye}, {Pierel}, {Chen}, {Kelly}, {Cohen}, {Windhorst}, {Riess}, {Kamieneski}, {Diego}, {Meena}, {Cha}, {Oguri}, {Zitrin}, {Jee}, {Foo}, {Leimbach}, {Koekemoer}, {Conselice}, {Dai}, {Goobar}, {Siebert}, {Strolger}, \& {Willner}}]{pascale+25}
{Pascale}, M., {Frye}, B.~L., {Pierel}, J. D.~R., {et~al.} 2025, \apj, 979, 13

\bibitem[{{Pierel} {et~al.}(2023){Pierel}, {Arendse}, {Ertl}, {Huang}, {Moustakas}, {Schuldt}, {Shajib}, {Shu}, {Birrer}, {Bronikowski}, {Hjorth}, {Suyu}, {Agarwal}, {Agnello}, {Bolton}, {Chakrabarti}, {Cold}, {Courbin}, {Della Costa}, {Dhawan}, {Engesser}, {Fox}, {Gall}, {Gomez}, {Goobar}, {Jha}, {Jimenez}, {Johansson}, {Larison}, {Li}, {Marques-Chaves}, {Mao}, {Mazzali}, {Perez-Fournon}, {Petrushevska}, {Poidevin}, {Rest}, {Sheu}, {Shirley}, {Silver}, {Storfer}, {Strolger}, {Treu}, {Wojtak}, \& {Zenati}}]{pierel23}
{Pierel}, J.~D.~R., {Arendse}, N., {Ertl}, S., {et~al.} 2023, \apj, 948, 115

\bibitem[{{Pierel} {et~al.}(2024){Pierel}, {Frye}, {Pascale}, {Caminha}, {Chen}, {Dhawan}, {Gilman}, {Grayling}, {Huber}, {Kelly}, {Thorp}, {Arendse}, {Birrer}, {Bronikowski}, {Ca{\~n}ameras}, {Coe}, {Cohen}, {Conselice}, {Driver}, {D{\'S}ilva}, {Engesser}, {Foo}, {Gall}, {Garuda}, {Grillo}, {Grogin}, {Henderson}, {Hjorth}, {Jansen}, {Johansson}, {Kamieneski}, {Koekemoer}, {Larison}, {Marshall}, {Moustakas}, {Nonino}, {Ortiz}, {Petrushevska}, {Pirzkal}, {Robotham}, {Ryan}, {Schuldt}, {Strolger}, {Summers}, {Suyu}, {Treu}, {Willmer}, {Windhorst}, {Yan}, {Zitrin}, {Acebron}, {Chakrabarti}, {Coulter}, {Fox}, {Huang}, {Jha}, {Li}, {Mazzali}, {Meena}, {P{\'e}rez-Fournon}, {Poidevin}, {Rest}, \& {Riess}}]{2024Pierel}
{Pierel}, J.~D.~R., {Frye}, B.~L., {Pascale}, M., {et~al.} 2024, \apj, 967, 50

\bibitem[{Pierel {et~al.}(2025)Pierel, Hayes, Millon, Larison, Mamuzic, Acebron, Agrawal, Bergamini, Cha, Dhawan, Diego, Frye, Gilman, Granata, Grillo, Jee, Kamieneski, Koekemoer, Meena, Newman, Oguri, Padilla-Gonzalez, Poidevin, Rosati, Schuldt, Strolger, Suyu, Thorp, \& Zitrin}]{pierel25}
Pierel, J. D.~R., Hayes, E.~E., Millon, M., {et~al.} 2025, submitted to A\&A [\eprint[arXiv]{2509.12301}]

\bibitem[{{Planck Collaboration} {et~al.}(2020){Planck Collaboration}, {Aghanim}, {Akrami}, {Ashdown}, {Aumont}, {Baccigalupi}, {Ballardini}, {Banday}, {Barreiro}, {Bartolo}, {Basak}, {Battye}, {Benabed}, {Bernard}, {Bersanelli}, {Bielewicz}, {Bock}, {Bond}, {Borrill}, {Bouchet}, {Boulanger}, {Bucher}, {Burigana}, {Butler}, {Calabrese}, {Cardoso}, {Carron}, {Challinor}, {Chiang}, {Chluba}, {Colombo}, {Combet}, {Contreras}, {Crill}, {Cuttaia}, {de Bernardis}, {de Zotti}, {Delabrouille}, {Delouis}, {Di Valentino}, {Diego}, {Dor{\'e}}, {Douspis}, {Ducout}, {Dupac}, {Dusini}, {Efstathiou}, {Elsner}, {En{\ss}lin}, {Eriksen}, {Fantaye}, {Farhang}, {Fergusson}, {Fernandez-Cobos}, {Finelli}, {Forastieri}, {Frailis}, {Fraisse}, {Franceschi}, {Frolov}, {Galeotta}, {Galli}, {Ganga}, {G{\'e}nova-Santos}, {Gerbino}, {Ghosh}, {Gonz{\'a}lez-Nuevo}, {G{\'o}rski}, {Gratton}, {Gruppuso}, {Gudmundsson}, {Hamann}, {Handley}, {Hansen}, {Herranz}, {Hildebrandt}, {Hivon}, {Huang}, {Jaffe}, {Jones}, {Karakci}, {Keih{\"a}nen},
  {Keskitalo}, {Kiiveri}, {Kim}, {Kisner}, {Knox}, {Krachmalnicoff}, {Kunz}, {Kurki-Suonio}, {Lagache}, {Lamarre}, {Lasenby}, {Lattanzi}, {Lawrence}, {Le Jeune}, {Lemos}, {Lesgourgues}, {Levrier}, {Lewis}, {Liguori}, {Lilje}, {Lilley}, {Lindholm}, {L{\'o}pez-Caniego}, {Lubin}, {Ma}, {Mac{\'\i}as-P{\'e}rez}, {Maggio}, {Maino}, {Mandolesi}, {Mangilli}, {Marcos-Caballero}, {Maris}, {Martin}, {Martinelli}, {Mart{\'\i}nez-Gonz{\'a}lez}, {Matarrese}, {Mauri}, {McEwen}, {Meinhold}, {Melchiorri}, {Mennella}, {Migliaccio}, {Millea}, {Mitra}, {Miville-Desch{\^e}nes}, {Molinari}, {Montier}, {Morgante}, {Moss}, {Natoli}, {N{\o}rgaard-Nielsen}, {Pagano}, {Paoletti}, {Partridge}, {Patanchon}, {Peiris}, {Perrotta}, {Pettorino}, {Piacentini}, {Polastri}, {Polenta}, {Puget}, {Rachen}, {Reinecke}, {Remazeilles}, {Renzi}, {Rocha}, {Rosset}, {Roudier}, {Rubi{\~n}o-Mart{\'\i}n}, {Ruiz-Granados}, {Salvati}, {Sandri}, {Savelainen}, {Scott}, {Shellard}, {Sirignano}, {Sirri}, {Spencer}, {Sunyaev}, {Suur-Uski}, {Tauber}, {Tavagnacco},
  {Tenti}, {Toffolatti}, {Tomasi}, {Trombetti}, {Valenziano}, {Valiviita}, {Van Tent}, {Vibert}, {Vielva}, {Villa}, {Vittorio}, {Wandelt}, {Wehus}, {White}, {White}, {Zacchei}, \& {Zonca}}]{planck20}
{Planck Collaboration}, {Aghanim}, N., {Akrami}, Y., {et~al.} 2020, \aap, 641, A6

\bibitem[{{Prajs} {et~al.}(2017){Prajs}, {Sullivan}, {Smith}, {Levan}, {Karpenka}, {Edwards}, {Walker}, {Wolf}, {Balland}, {Carlberg}, {Howell}, {Lidman}, {Pain}, {Pritchet}, \& {Ruhlmann-Kleider}}]{Prajs2017}
{Prajs}, S., {Sullivan}, M., {Smith}, M., {et~al.} 2017, \mnras, 464, 3568

\bibitem[{{Quimby} {et~al.}(2011){Quimby}, {Kulkarni}, {Kasliwal}, {Gal-Yam}, {Arcavi}, {Sullivan}, {Nugent}, {Thomas}, {Howell}, {Nakar}, {Bildsten}, {Theissen}, {Law}, {Dekany}, {Rahmer}, {Hale}, {Smith}, {Ofek}, {Zolkower}, {Velur}, {Walters}, {Henning}, {Bui}, {McKenna}, {Poznanski}, {Cenko}, \& {Levitan}}]{quimby2011}
{Quimby}, R.~M., {Kulkarni}, S.~R., {Kasliwal}, M.~M., {et~al.} 2011, \nat, 474, 487

\bibitem[{{Refsdal}(1964)}]{1964Refsdal}
{Refsdal}, S. 1964, \mnras, 128, 307

\bibitem[{{Riello} {et~al.}(2021){Riello}, {De Angeli}, {Evans}, {Montegriffo}, {Carrasco}, {Busso}, {Palaversa}, {Burgess}, {Diener}, {Davidson}, {Rowell}, {Fabricius}, {Jordi}, {Bellazzini}, {Pancino}, {Harrison}, {Cacciari}, {van Leeuwen}, {Hambly}, {Hodgkin}, {Osborne}, {Altavilla}, {Barstow}, {Brown}, {Castellani}, {Cowell}, {De Luise}, {Gilmore}, {Giuffrida}, {Hidalgo}, {Holland}, {Marinoni}, {Pagani}, {Piersimoni}, {Pulone}, {Ragaini}, {Rainer}, {Richards}, {Sanna}, {Walton}, {Weiler}, \& {Yoldas}}]{2021A&A...649A...3R}
{Riello}, M., {De Angeli}, F., {Evans}, D.~W., {et~al.} 2021, \aap, 649, A3

\bibitem[{{Riess} {et~al.}(2022){Riess}, {Yuan}, {Macri}, {Scolnic}, {Brout}, {Casertano}, {Jones}, {Murakami}, {Anand}, {Breuval}, {Brink}, {Filippenko}, {Hoffmann}, {Jha}, {D'arcy Kenworthy}, {Mackenty}, {Stahl}, \& {Zheng}}]{2022riess}
{Riess}, A.~G., {Yuan}, W., {Macri}, L.~M., {et~al.} 2022, \apjl, 934, L7

\bibitem[{{Rodney} {et~al.}(2021){Rodney}, {Brammer}, {Pierel}, {Richard}, {Toft}, {O'Connor}, {Akhshik}, \& {Whitaker}}]{rodney21}
{Rodney}, S.~A., {Brammer}, G.~B., {Pierel}, J. D.~R., {et~al.} 2021, Nature Astronomy [\eprint[arXiv]{2106.08935}]

\bibitem[{{Sainz de Murieta} {et~al.}(2024){Sainz de Murieta}, {Collett}, {Magee}, {Pierel}, {Enzi}, {Lokken}, {Gagliano}, \& {Ryczanowski}}]{SainzdeMurieta+24}
{Sainz de Murieta}, A., {Collett}, T.~E., {Magee}, M.~R., {et~al.} 2024, \mnras, 535, 2523

\bibitem[{Smith {et~al.}(2019)Smith, Williams, Young, Ibsen, Smartt, Lawrence, Morris, Voutsinas, \& Nicholl}]{smith19}
Smith, K.~W., Williams, R.~D., Young, D.~R., {et~al.} 2019, Research Notes of the {AAS}, 3, 26

\bibitem[{{Smith} {et~al.}(2018){Smith}, {Sullivan}, {Nichol}, {Galbany}, {D'Andrea}, {Inserra}, {Lidman}, {Rest}, {Schirmer}, {Filippenko}, {Zheng}, {Cenko}, {Angus}, {Brown}, {Davis}, {Finley}, {Foley}, {Gonz{\'a}lez-Gait{\'a}n}, {Guti{\'e}rrez}, {Kessler}, {Kuhlmann}, {Marriner}, {M{\"o}ller}, {Nugent}, {Prajs}, {Thomas}, {Wolf}, {Zenteno}, {Abbott}, {Abdalla}, {Allam}, {Annis}, {Bechtol}, {Benoit-L{\'e}vy}, {Bertin}, {Brooks}, {Burke}, {Carnero Rosell}, {Carrasco Kind}, {Carretero}, {Castander}, {Crocce}, {Cunha}, {da Costa}, {Davis}, {Desai}, {Diehl}, {Doel}, {Eifler}, {Flaugher}, {Fosalba}, {Frieman}, {Garc{\'\i}a-Bellido}, {Gaztanaga}, {Gerdes}, {Goldstein}, {Gruen}, {Gruendl}, {Gschwend}, {Gutierrez}, {Honscheid}, {James}, {Johnson}, {Kuehn}, {Kuropatkin}, {Li}, {Lima}, {Maia}, {Marshall}, {Martini}, {Menanteau}, {Miller}, {Miquel}, {Ogando}, {Petravick}, {Plazas}, {Romer}, {Rykoff}, {Sako}, {Sanchez}, {Scarpine}, {Schindler}, {Schubnell}, {Sevilla-Noarbe}, {Smith}, {Soares-Santos}, {Sobreira},
  {Suchyta}, {Swanson}, {Tarle}, {Walker}, \& {DES Collaboration}}]{smith2018}
{Smith}, M., {Sullivan}, M., {Nichol}, R.~C., {et~al.} 2018, \apj, 854, 37

\bibitem[{Suyu {et~al.}(2025)Suyu, Acebron, Grillo, Bergamini, Caminha, Cha, Diego, Ertl, Foo, Frye, Fudamoto, Granata, Halkola, Jee, Kamieneski, Koekemoer, Meena, Newman, Nishida, Oguri, Rosati, Schuldt, Zitrin, Cañameras, Hayes, Larison, Mamuzic, Millon, Pierel, Tortorelli, \& Wang}]{suyu25}
Suyu, S.~H., Acebron, A., Grillo, C., {et~al.} 2025, submitted to A\&A [\eprint[arXiv]{2509.12319}]

\bibitem[{{Suyu} {et~al.}(2017){Suyu}, {Bonvin}, {Courbin}, {Fassnacht}, {Rusu}, {Sluse}, {Treu}, {Wong}, {Auger}, {Ding}, {Hilbert}, {Marshall}, {Rumbaugh}, {Sonnenfeld}, {Tewes}, {Tihhonova}, {Agnello}, {Blandford}, {Chen}, {Collett}, {Koopmans}, {Liao}, {Meylan}, \& {Spiniello}}]{Suyu17}
{Suyu}, S.~H., {Bonvin}, V., {Courbin}, F., {et~al.} 2017, \mnras, 468, 2590

\bibitem[{{Suyu} {et~al.}(2024){Suyu}, {Goobar}, {Collett}, {More}, \& {Vernardos}}]{Suyu+24}
{Suyu}, S.~H., {Goobar}, A., {Collett}, T., {More}, A., \& {Vernardos}, G. 2024, \ssr, 220, 13

\bibitem[{{Suyu} {et~al.}(2020){Suyu}, {Huber}, {Ca{\~n}ameras}, {Kromer}, {Schuldt}, {Taubenberger}, {Y{\i}ld{\i}r{\i}m}, {Bonvin}, {Chan}, {Courbin}, {N{\"o}bauer}, {Sim}, \& {Sluse}}]{Suyu+20}
{Suyu}, S.~H., {Huber}, S., {Ca{\~n}ameras}, R., {et~al.} 2020, \aap, 644, A162

\bibitem[{{Suyu} {et~al.}(2010){Suyu}, {Marshall}, {Auger}, {Hilbert}, {Blandford}, {Koopmans}, {Fassnacht}, \& {Treu}}]{suyu10b}
{Suyu}, S.~H., {Marshall}, P.~J., {Auger}, M.~W., {et~al.} 2010, \apj, 711, 201

\bibitem[{{TDCOSMO Collaboration} {et~al.}(2025){TDCOSMO Collaboration}, {Birrer}, {Buckley-Geer}, {Cappellari}, {Courbin}, {Dux}, {Fassnacht}, {Frieman}, {Galan}, {Gilman}, {Huang}, {Knabel}, {Langeroodi}, {Lin}, {Millon}, {Morishita}, {Motta}, {Mozumdar}, {Paic}, {Shajib}, {Sheu}, {Sluse}, {Sonnenfeld}, {Spiniello}, {Stiavelli}, {Suyu}, {Tan}, {Treu}, {Van de Vyvere}, {Wang}, {Wells}, {Williams}, \& {Wong}}]{TDCOSMO2025}
{TDCOSMO Collaboration}, {Birrer}, S., {Buckley-Geer}, E.~J., {et~al.} 2025, arXiv e-prints, arXiv:2506.03023

\bibitem[{{Treu} {et~al.}(2022){Treu}, {Suyu}, \& {Marshall}}]{Treu+22}
{Treu}, T., {Suyu}, S.~H., \& {Marshall}, P.~J. 2022, \aapr, 30, 8

\bibitem[{{Vanzella} {et~al.}(2017){Vanzella}, {Calura}, {Meneghetti}, {Mercurio}, {Castellano}, {Caminha}, {Balestra}, {Rosati}, {Tozzi}, {De Barros}, {Grazian}, {D'Ercole}, {Ciotti}, {Caputi}, {Grillo}, {Merlin}, {Pentericci}, {Fontana}, {Cristiani}, \& {Coe}}]{vanzella17}
{Vanzella}, E., {Calura}, F., {Meneghetti}, M., {et~al.} 2017, \mnras, 467, 4304

\bibitem[{{Verde} {et~al.}(2024){Verde}, {Sch{\"o}neberg}, \& {Gil-Mar{\'\i}n}}]{Verde+24}
{Verde}, L., {Sch{\"o}neberg}, N., \& {Gil-Mar{\'\i}n}, H. 2024, \araa, 62, 287

\bibitem[{{Wang} {et~al.}(2023){Wang}, {Liu}, {Cai}, {Geng}, {Fang}, {He}, {Jiang}, {Jiang}, {Kong}, {Li}, {Li}, {Luo}, {Pan}, {Wu}, {Yang}, {Yu}, {Zheng}, {Zhu}, {Cai}, {Chen}, {Chen}, {Dai}, {Fan}, {Fan}, {Fang}, {He}, {Hu}, {Hu}, {Jin}, {Jiang}, {Li}, {Li}, {Li}, {Liang}, {Lin}, {Liu}, {Liu}, {Liu}, {Liu}, {Liu}, {Lou}, {Qu}, {Sheng}, {Shi}, {Shu}, {Su}, {Sun}, {Wang}, {Wang}, {Wang}, {Wang}, {Wei}, {Wei}, {Xue}, {Yan}, {Yang}, {Yuan}, {Yuan}, {Zhang}, {Zhang}, {Zhao}, \& {Zhao}}]{WFST23}
{Wang}, T., {Liu}, G., {Cai}, Z., {et~al.} 2023, Science China Physics, Mechanics, and Astronomy, 66, 109512

\bibitem[{{Wise} {et~al.}(2025){Wise}, {Perley}, {Goobar}, {Johansson}, \& {McGrath}}]{2025TNSconf}
{Wise}, J., {Perley}, D., {Goobar}, A., {Johansson}, J., \& {McGrath}, Z. 2025, Transient Name Server AstroNote, 296, 1

\bibitem[{{Wojtak} {et~al.}(2019){Wojtak}, {Hjorth}, \& {Gall}}]{wojtak2019}
{Wojtak}, R., {Hjorth}, J., \& {Gall}, C. 2019, \mnras, 487, 3342

\bibitem[{{Wong} {et~al.}(2020){Wong}, {Suyu}, {Chen}, {Rusu}, {Millon}, {Sluse}, {Bonvin}, {Fassnacht}, {Taubenberger}, {Auger}, {Birrer}, {Chan}, {Courbin}, {Hilbert}, {Tihhonova}, {Treu}, {Agnello}, {Ding}, {Jee}, {Komatsu}, {Shajib}, {Sonnenfeld}, {Bland ford}, {Koopmans}, {Marshall}, \& {Meylan}}]{Wong20}
{Wong}, K.~C., {Suyu}, S.~H., {Chen}, G. C.~F., {et~al.} 2020, \mnras [\eprint[arXiv]{1907.04869}]

\bibitem[{{Yan} {et~al.}(2017){Yan}, {Quimby}, {Gal-Yam}, {Brown}, {Blagorodnova}, {Ofek}, {Lunnan}, {Cooke}, {Cenko}, {Jencson}, \& {Kasliwal}}]{yan2017}
{Yan}, L., {Quimby}, R., {Gal-Yam}, A., {et~al.} 2017, \apj, 840, 57

\bibitem[{{Zhou} {et~al.}(2023){Zhou}, {Ferraro}, {White}, {DeRose}, {Sailer}, {Aguilar}, {Ahlen}, {Bailey}, {Brooks}, {Claybaugh}, {Dawson}, {de la Macorra}, {Dey}, {Doel}, {Font-Ribera}, {Forero-Romero}, {Gontcho A Gontcho}, {Guy}, {Kremin}, {Lambert}, {Le Guillou}, {Levi}, {Magneville}, {Manera}, {Meisner}, {Miquel}, {Moustakas}, {Myers}, {Newman}, {Nie}, {Percival}, {Rezaie}, {Rossi}, {Sanchez}, {Schlegel}, {Schubnell}, {Seo}, {Tarl{\'e}}, \& {Zhou}}]{zhou23}
{Zhou}, R., {Ferraro}, S., {White}, M., {et~al.} 2023, \jcap, 2023, 097

\end{thebibliography}


\end{document}